\documentclass[10pt,final,twocolumn,twoside,journal,romanappendices]{IEEEtran}
\usepackage{cite}
\usepackage{epstopdf}
\usepackage{epsfig}
\usepackage{amsthm}
\usepackage{amsmath,amssymb,amsfonts,amstext,amsbsy,amsopn,dsfont}
\usepackage{cases}
\usepackage{stfloats}
\usepackage{color}
\usepackage{sublabel}
\usepackage{bigints}
\usepackage{array}

\newtheorem{proposition}{\textbf{Proposition}}

\newtheorem{remark}{\textbf{Remark}}

\newcommand{\defn}{\triangleq}
\newcommand{\dif}{\mathrm{d}}

\definecolor{barered}{rgb}{1.0, 0.0, 0.0}

\begin{document}

\title{Energy-Efficient Activation and Uplink Transmission for Cellular IoT}

\author{Chun-Hung Liu, Yu-Han Shen, and Chia-Han Lee\\
	\thanks{C.-H. Liu is with the Department of Electrical and Computer Engineering at Mississippi State University, Mississippi State 39762, USA. Y.-H. Shen and C.-H. Lee are with the Institute of Communications Engineering and Department of Electrical and Computer Engineering at National Chiao Tung University, Hsinchu 30010, Taiwan. The corresponding author is Dr. Liu  (e-mail: chliu@ece.msstate.edu).}
}


\maketitle

\begin{abstract}
Consider a large-scale cellular network in which base stations (BSs) serve massive Internet-of-Things (IoT) devices. Since IoT devices are powered by a capacity-limited battery, how to prolong their working lifetime is a paramount problem for the success of cellular IoT systems. This paper proposes how to use BSs to manage the active and dormant operating modes of the IoT devices via downlink signaling in an energy-efficient fashion and how the IoT devices perform energy-efficient uplink power control to improve their uplink coverage. We first investigate the fundamental statistical properties of an activation signaling process induced by BSs that would like to activate the devices in their cells, which helps to derive the neat expressions of the true, false and total activation probabilities that reveal joint downlink power control and BS coordination is an effective means to significantly improve the activation performance. We then propose an energy-efficient uplink power control for IoT devices which is shown to save power and ameliorate the uplink coverage probability at the same time. We also propose an energy-efficient downlink power control and BS coordination scheme, which is shown to remarkably improve the activation and uplink coverage performances at the same time.
\end{abstract}


\section{Introduction}
Connecting numerous physical things (such as sensors) to the Internet by having them piggyback on cellular networks is a promising way for Internet of Things (IoT) communications since cellular networks are able to provide a seamless coverage of mobile devices, low-cost multicasting and broadcasting services \cite{AFMGMM15}. For example, Long Term Evolution for Machines (LTE-M) is a network standard that enables (IoT) devices to piggyback on existing cellular networks so that the devices with a suitable software update can communicate with the cloud over cellular base stations (BSs). Cellular IoT devices are particularly suitable for \textit{mission-critical} applications in which real-time data transfer makes the difference, e.g., self-driving cars or emergency devices in smart cities \cite{AZNBAC14,HLQLDT19,FQXZGM19}. In addition, densely deploying different kinds of BSs in a heterogeneous cellular network significantly reduces the link budget between devices and their serving BSs so that the devices can use low transmit power for uplink transmission. 

Despise these aforementioned advantages of cellular IoT, cellular networks face a serious problem while serving massive IoT devices operated with capacity-limited batteries.  How to make the devices operate in an energy-efficient way so as to prolong their working lifetime is critical. There are essentially two methods to extend the working lifetime of IoT devices. One method is to design low power-consuming and/or energy-harvesting circuits for IoT devices  \cite{XLNA19}\cite{AFTFOB19}. Although such a circuit design method may largely reduce the power consumption of IoT devices and may allow opportunistically replenish battery power, it incurs high hardware design cost. The other method is to let IoT devices switch between the ON (active) and OFF (dormant) operating modes according to traffic patterns to save power. Such a mode-switched method, in general, is of low complexity and low cost because we only need to design a simple ON-OFF signaling protocol to be implemented by the devices and their serving BSs. Most importantly, this mode-switched method may save much more power than the circuit design method because IoT devices usually have much less uplink traffic and they thus operate in the dormant mode for most of the time, whereas the circuit design method only helps when they are active. 
 
\subsection{Motivation and Prior Work}
The mode-switched method can be briefly described as follows. Each BS occasionally broadcasts activation signals to instruct the devices in its cell to switch to the active mode for uplink transmission. Once the activated devices finish the uplink transmission and do not receive activation signals after some time, they automatically switch to the dormant mode. This  mode-switched method motivates us to study the following two problems. The first one is how BSs broadcast their activation signals to correctly activate their serving devices. The second one is how the activated devices save their transmit power and improve their uplink transmission performance. Most of the prior works on energy-efficient communications in IoT networks do not address these two problems and mainly focus on power allocation, wireless power transfer and energy-harvesting problems of saving and replenishing the power of IoT devices (typically see \cite{ZCFZZZ18,KWCAAADS18,DZRZLC18,AOEMOSIFA18,MAKHSD18,WSJL18,DSGHHNHDT19,DSEETMD19,RJKXPF19,NKZDJGA18}). In reference \cite{KWCAAADS18}, for example, a distributed wireless power transfer system with or without the frequency and phase synchronization is proposed to charge IoT devices by using power beacons. Reference \cite{DZRZLC18} studied how to do energy-efficient dynamic user scheduling and power allocation by minimizing the total power consumption of a wireless network with massive IoT devices under the constraint on the long-term rate requirements of users. Reference \cite{MAKHSD18} jointly analyzed the downlink and uplink coverage problem of RF-powered IoT devices in a cellular network, but it did not address how to jointly perform downlink and uplink energy-efficient transmission based on the analytical coverage results. In reference \cite{RJKXPF19}, a power minimization problem of a simultaneous wireless information and power transfer network with coexisting power-splitting IoT devices and time-switching IoT devices was investigated under a nonlinear energy harvesting model. Although reference \cite{NKZDJGA18} studied the RF wake-up problem of IoT devices that is similar to the first problem mentioned above, it does not study how to control the downlink transmit power of BSs to save power and improve the probability of correctly activating the devices as much as possible.

\subsection{Contributions}
Since the aforementioned two problems are barely investigated in the literature, this paper is dedicated to thoroughly studying them and its contributions are summarized as follows:
\begin{itemize}
	\item An activation signaling process in a Poisson cellular network is proposed to characterize the total activation signal power received by a device. The false, true and total activation probabilities are defined based on the activation signaling process, and the Laplace transforms of the activation signaling process and its sub-processes are derived in a form that is easy to compute.
	\item We adopt the technique of inverse Laplace transform to explicitly find the expressions of the false, true and total activation probabilities for downlink channels with Nakagami-$m$ fading and show that they can be reduced to nearly a closed form when the path-loss exponent is equal to four. From these expressions, we learn that joint downlink power control and BS coordination can effectively increase the true activation probability and reduce the false activation probability at the same time and the activation performance is thus improved.
	\item An energy-efficient uplink power control scheme is proposed to save power for IoT devices. For the proposed scheme, we not only derive the uplink coverage (probability) for uplink channels with Nakagami-$m$ fading and find its closed form when the path-loss exponent is equal to four, but also show that the scheme is able to save power and improve the uplink coverage  at the same time if the system parameters are chosen appropriately. 
	\item We also propose an energy-efficient downlink power control and BS coordination scheme to save power as well as improve the activation performance. The false, true, and total activation probabilities are derived under the proposed scheme, and how the scheme is able to enhance the uplink coverage is shown. Finally, we propose an activation performance index to quantitatively evaluate the activation performance of the proposed scheme under various parameter settings, which can provide insights into how to choose the system parameters for the proposed scheme.
\end{itemize} 
Furthermore, numerical simulation results are provided to validate the correctness and accuracy of our analytical findings and support our observations and discussions. 

\subsection{Paper Organization}
The rest of this paper is organized as follows. In Section \ref{Sec:SystemModel}, the system model and its related assumptions are first specified and then the preliminaries regarding the activation signaling process are introduced. Section \ref{Sec:SuccActProb} analyzes the total activation probability and uplink coverage with energy-efficient power control. In Section \ref{Sec:EngEffPowConBScoor}, an energy-efficient downlink power control and BS coordination scheme is proposed, and how it affects the activation performance and uplink coverage is studied. Finally, Section \ref{Sec:Conclusion} concludes our analytical findings and observations.

\section{System Model and Preliminaries}\label{Sec:SystemModel}
Consider a large-scale planar cellular network in which all (small cell) BSs form an independent homogeneous Poisson point process (PPP) of density $\lambda_b$ and they can be expressed as the set $\Phi_b$ given by
\begin{align}
\Phi_b\defn \{X_i\in\mathbb{R}^2: i\in\mathbb{N}\},
\end{align} 
where $X_i$ denotes BS $i$ and its location. Suppose there are numerous (IoT) devices which are randomly and independently distributed over the entire cellular network. These IoT devices form an independent homogeneous PPP of density $\lambda_d$ and can be represented by the following set:
\begin{align*}
\Phi_d \defn\{D_j\in\mathbb{R}^2: j\in\mathbb{N}\},
\end{align*}
where $D_j$ denotes device $j$ and its location. All the devices are assumed to simply have two power operating modes similar to the modes proposed in \cite{NKZDJGA18}: one is the \textit{\textbf{active}} mode and the other is the \textit{\textbf{dormant}} mode. If the devices are not requested to perform uplink transmission by their serving BSs after some short period of time, they automatically switch to the dormant mode to save power, yet they remain registered in the network. Each BS occasionally sends a signal to activate (wake up) the dormant devices in its cell so as to switch them back to the active mode and requests them to perform uplink transmission. We assume all the IoT devices are served by their nearest BS and all the BSs in the network use the same radio resource (channel) to manage the power operating modes of the IoT devices. According to Lemma 2 in our previous work \cite{CHLHCT17}, such a nearest BS association scheme gives rise to the following probability mass function (PMF) of the number of the devises associated with a BS:
\begin{align}\label{Eqn:DisNumDev}
\phi_n \defn \frac{\Gamma(n+\frac{7}{2})}{n! \Gamma(\frac{7}{2})}\left(1+\frac{2\lambda_d}{7\lambda_b}\right)^{-(n+\frac{7}{2})},
\end{align}
where $\phi_n$ is the probability that there are $n$ devices associated with a BS and $\Gamma(x)\defn \int_0^{\infty} t^{x-1}e^{-t}\dif t$ is the Gamma function. The density ratio of $\lambda_d/\lambda_b$ in \eqref{Eqn:DisNumDev} denotes the average number of the devices in a cell since all the cells of the BSs in the network are Voroni-tessellated and each of them has an average area of $1/\lambda_b$ \cite{DSWKJM13}\cite{FBBBL10}. Note that the density $\lambda_d$ of the devices is much larger than that of the BSs in practice so that each BS almost surely associates with at least one device and thereby almost no void cells exist in the network \cite{CHLLCW1502,CHLLCW16}, i.e., $\phi_0=(1+2\lambda_d/7\lambda_b)^{-\frac{7}{2}}\approx 0$ owing to $\frac{\lambda_d}{\lambda_b} \gg 1$. Thus, the void cell phenomenon due to BS association will not be considered in the following modeling and analysis.

\subsection{Channel Model and activation signaling process}
Suppose there is a typical device located at the origin and it is associated with BS $X_1\in\Phi_b$ that is the nearest BS to it.  The channel between a BS and a device undergoes path loss and small-scale fading. Let the transmit power of a BS be constant $\overline{P}$, and then the total aggregated power of the activation signals from all the BSs received by the typical device can be written as\footnote{According to the Slivnyak theorem \cite{FBBBL10,MHRKG09,DSWKJM13}, the statistical properties evaluated at the origin are the same as those evaluated at any particular point in a PPP. As such, many of the following equations and derivations are expressed and analyzed based on the origin, the location of the typical device. }
\begin{align}\label{Eqn:ShotWakeUpPower}
W\defn \omega_1\underbrace{\overline{P}H_1\|X_1\|^{-\alpha}}_{D_1}+\underbrace{\sum_{i:X_i\in\Phi_{b,1}}\omega_i\overline{P}H_i\|X_i\|^{-\alpha}}_{I_1},
\end{align}
where $\omega_i\in\{0,1\}$ is a Bernoulli random variable (RV) that is unity if BS $X_i$ is transmitting its activation signal and zero otherwise, $H_i$ is the fading channel gain from BS $X_i$ to the typical device, $\|Y_i-Y_j\|$ denotes the Euclidean distance between nodes $Y_i$ and $Y_j$ for $i\neq j$, $\Phi_{b,1}\defn \Phi_b\setminus X_1$ is the set of the BSs without $X_1$, and $\alpha>2$ denotes the path-loss exponent. All the small-scale fading channel gains in \eqref{Eqn:ShotWakeUpPower} are assumed to be i.i.d. and they can be characterized by the Nakagami-$m$ fading model\footnote{The Nakagami-$m$ fading model is adopted in this paper because it covers several different fading models, such as Rayleigh fading (for $m=1$), Rician fading with parameter $L$ (for $m=\frac{(L+1)^2}{2L+1}$), no fading (for $m\rightarrow\infty$), and other intermediate fading distributions. In addition, for the tractable analysis in the sequel, $m$ is assumed as a non-negative integer if needed.} so that $H_i\sim\mathrm{Gamma}(m,m)$ is a Gamma RV with shape $m$ and rate $m$ for all $i\in\mathbb{N}_+$. In this paper, we call $W$ \textit{activation (shot) signaling process} because it captures the cumulative effect at a device of a set of random shock signals appearing at random locations $X_i$, and $\omega_i\overline{P}H_i\|X_i\|^{-\alpha}$ can be viewed as the impulse function that gives the attenuation of the transmit power of a BS in space \cite{SBLMCT90}. Note that $D_1$ and $I_1$ in \eqref{Eqn:ShotWakeUpPower} represent the power of the \textit{desired} activation signal from $X_1$ and the aggregated power of all the \textit{non-desired} interfering activation signals from all the BSs in set  $\Phi_{b,1}$, respectively. When BS $X_1$ broadcasts an activation signal in its cell, $D_1$ and $I_1$ both exist in \eqref{Eqn:ShotWakeUpPower} so that $I_1$ is able to help $X_1$ activate its devices in this case. When BS $X_1$ remains idle, only $I_1$ exists in \eqref{Eqn:ShotWakeUpPower} and the devices served by X1 may be accidentally activated if $I_1$ is sufficiently large. This indicates that $I_1$ plays a crucial role in activating the devices in a cell and we need to appropriately exploit it so as to boost the activation performance.

\subsection{The Statistical Properties of activation signaling process $W$}

According to the activation signaling process in \eqref{Eqn:ShotWakeUpPower}, the total activation probability of a device can be defined as follows:
\begin{align}\label{Eqn:DefActProb}
\eta_a\defn \mathbb{P}[W\geq \theta_a] =& \mathbb{P}[W\geq\theta_a|\omega_1=1]\mathbb{P}[\omega_1=1]\nonumber\\
&+\mathbb{P}[W\geq\theta_a|\omega_1=0]\mathbb{P}[\omega_1=0]\nonumber\\
=&\mu p_a + (1-\mu) q_a=q_a+\mu (p_a-q_a),
\end{align}
where $\theta_a$ denotes the activation threshold that is the minimum power required to activate a device, $\mu\defn \mathbb{P}[\omega_i=1]$ (and thus $\mathbb{P}[\omega_i=0]=1-\mu$) for all $i\in\mathbb{N}_+$, $p_a\defn \mathbb{P}[W\geq\theta_a|\omega_1=1]$ is called the \textit{true activation probability}, and $q_a\defn \mathbb{P}[W\geq\theta_a|\omega_1=0]$ is called the \textit{false activation probability}. Note that $p_a$ reveals how likely a device is successfully activated by the BS associating with it whereas $q_a$ reflects how likely a device is accidentally activated by the BSs not associating with it. Moreover, $q_a$ also indicates whether a device in the network operates in an energy-efficient status in that a small value of $q_a$ implies that devices do not loose much energy due to false activation. As such, an active BS has a high activation performance if it can attain a high true activation probability and a low false activation probability while activating its devices.

To achieve the goal of high activation performance, we need to increase $p_a$ and lower $q_a$ as much as possible. Generally exploiting the statistical properties of the activation signaling process $W$ in \eqref{Eqn:ShotWakeUpPower} sheds light on how to efficiently achieve this goal. In particular, we are interested in the Laplace transforms of $W$, $I_1$ and $D_1+I_1$ because they will facilitate our following analyses and derivations regarding the total activation probability of a device. The Laplace transform of a non-negative RV $Z$ is defined as $\mathcal{L}_{Z}(s)\defn \mathbb{E}[\exp(-sZ)]$ for $s>0$ and thereby $\mathcal{L}_W(s)$, $\mathcal{L}_{I_1}(s)$ and $\mathcal{L}_{D_1+I_1}(s)$ are explicitly found as shown in the following proposition.
\begin{proposition}\label{Prop:LapWakeUpSignal}
For the activation signaling process $W$ in \eqref{Eqn:ShotWakeUpPower}, its Laplace transform is found as
\begin{align}\label{Eqn:LapTranWakeUpProc}
\mathcal{L}_{W}(s) &= \exp\left\{-\frac{\pi\lambda_b\mu (s\overline{P})^{\frac{2}{\alpha}}\Gamma(m+\frac{2}{\alpha})\Gamma(1-\frac{2}{\alpha})}{\Gamma(m)}\right\}\nonumber\\
&=\exp\left\{-\pi\lambda_b\mu\mathfrak{I}(0,s\overline{P})\right\},
\end{align}
where $\mathfrak{I}(\cdot,\cdot)$ is defined as
\begin{align}\label{Eqn:FunInterference}
\mathfrak{I}(x,y) \defn& y^{\frac{2}{\alpha}}\frac{\Gamma(m+\frac{2}{\alpha})\Gamma(1-\frac{2}{\alpha})}{\Gamma(m )}\nonumber\\
&-y^{\frac{2}{\alpha}}\int_0^{y^{-\frac{2}{\alpha}}x}\left[1-\left(\frac{t^{\frac{\alpha}{2}}}{t^{\frac{\alpha}{2}}+m}\right)^m\right]\dif t.
\end{align}
The Laplace transform of $W$ for given $\omega_1=0$, i.e., $\mathbb{E}[\exp(-sW)|\omega_1=0]= \mathcal{L}_{I_1}(s)$, can be found as
\begin{align}\label{Eqn:CondLapWd1}
\mathcal{L}_{I_1}(s) =\int_0^{\infty} \pi\lambda_b \exp\left\{-\pi\lambda_b \left[x+\mu \mathfrak{I}\left(x,s\overline{P}\right) \right]\right\}\dif x.
\end{align}
By using \eqref{Eqn:LapTranWakeUpProc} and \eqref{Eqn:CondLapWd1}, the Laplace transform of $W$ for given $\omega_1=1$, i.e., $\mathbb{E}[\exp(-sW)|\omega_1=1]\defn \mathcal{L}_{D_1+I_1}(s)$, can be obtained as
\begin{align}\label{Eqn:CondLapWd2}
\mathcal{L}_{D_1+I_1}(s) = \frac{1}{\mu}\left[\mathcal{L}_{W}(s)-(1-\mu)\mathcal{L}_{I_1}(s) \right].
\end{align}
\end{proposition}
\begin{IEEEproof}
See Appendix \ref{App:LapTranWakeUpProc}.
\end{IEEEproof}
\begin{remark}\label{Rem:SpeFunInterference}
If the integral in the second term of \eqref{Eqn:FunInterference} has $y=s\overline{P}$, then the mean value theorem in Calculus indicates that there exists an $x_1\in[0, (s\overline{P})^{-\frac{2}{\alpha}}x]$ for any $x>0$ such that the following identity
\begin{align*}
&(s\overline{P})^{\frac{2}{\alpha}}\int_0^{(s\overline{P})^{-\frac{2}{\alpha}}x} \left[1-\left(\frac{t^{\frac{\alpha}{2}}}{t^{\frac{\alpha}{2}}+m}\right)^m\right]\dif t\\
&=\left(1-\left[\frac{x^{\frac{\alpha}{2}}_1}{x^{\frac{\alpha}{2}}_1+m}\right]^m\right) x\defn \epsilon_1 x
\end{align*}
holds for $\epsilon_1\defn1-\left[\frac{x^{\frac{\alpha}{2}}_1}{x^{\frac{\alpha}{2}}_1+m}\right]^m$. Thus, there exists an $\epsilon_1\in[0,1)$ such that $\mathfrak{I}(x,s)$ in \eqref{Eqn:FunInterference} can be equivalently written as
\begin{align}
\mathfrak{I}(x,s\overline{P})  &=(s\overline{P})^{\frac{2}{\alpha}}\frac{\Gamma(m+\frac{2}{\alpha})\Gamma(1-\frac{2}{\alpha})}{\Gamma(m)}-\epsilon_1 x\nonumber\\
&=\mathfrak{I}(0,s\overline{P})-\epsilon_1 x. \label{Eqn:IdenJfun}
\end{align}
In light of this, if $\epsilon_1$ is not sensitive to $x$ (e.g., $m\gg x_1$ and $m \ll x_1$), then \eqref{Eqn:CondLapWd1} approximately reduces to
\begin{align}\label{Eqn:LapTransIntFirst}
\mathcal{L}_{I_1}(s) \approx \frac{\exp\left[-\pi\lambda_b\mu\mathfrak{I}(0,s\overline{P})\right]}{1-\mu\epsilon_1}.
\end{align}
Such an approximation is very accurate, which will be numerically demonstrated in Section \ref{SubSec:NumSuccActProb}.
\end{remark}
\noindent Note that the closed-form result in \eqref{Eqn:LapTranWakeUpProc} is essentially similar to the Laplace transform of the interference of a Poisson wireless ad hoc network in the literature (e.g., see \cite{FBBBL10,MHRKG09}), but it is much more general. Also note that the result in \eqref{Eqn:CondLapWd1} is similar to the result of Theorem 1 in our previous work \cite{CHLCSH19}, whereas the result in \eqref{Eqn:CondLapWd2} has not been found in the literature. Although the results in Proposition \ref{Prop:ActProb} are not completely found in closed form, they are able to reduce to very neat expressions for some special cases. For example, if $s=x^{\frac{\alpha}{2}}/\overline{P}$ and considering Rayleigh fading (i.e., $m=1$), then $\mathcal{L}_{I_1}(s)$ in \eqref{Eqn:CondLapWd1} exactly reduces to
\begin{align}\label{Eqn:LapTranFalseActProcSpec}
\mathcal{L}_{I_1}\left(\frac{x^{\frac{\alpha}{2}}}{\overline{P}}\right) =  \frac{1}{1+\mu \mathfrak{I}(1,1)},
\end{align}
where $\mathfrak{I}(1,1)$ is given by
\begin{align*}
\mathfrak{I}(1,1) = \frac{1}{\mathrm{sinc}(2/\alpha)}-\int_0^1 \frac{\dif t}{1+t^{\frac{\alpha}{2}}}
\end{align*}
in which $\mathrm{sinc}(2/\alpha)\defn \frac{\sin(2\pi/\alpha)}{2/\alpha}=1/\Gamma(1+\frac{2}{\alpha})\Gamma(1-\frac{2}{\alpha})$
and $\mathcal{L}_{D_1+I_1}(s)$ in \eqref{Eqn:CondLapWd2} can be simply expressed as
\begin{align}\label{Eqn:LapTranTrueActProcSpec}
\mathcal{L}_{D_1+I_1}\left(\frac{x^{\frac{\alpha}{2}}}{\overline{P}}\right)  = \frac{1}{\mu}\left\{\frac{\mathrm{sinc}(2/\alpha)}{\mathrm{sinc}(2/\alpha)+\mu}-\frac{(1-\mu)}{1+\mu \mathfrak{I}(1,1)}\right\}.
\end{align}
Note that the results in \eqref{Eqn:LapTranFalseActProcSpec} and \eqref{Eqn:LapTranTrueActProcSpec} do not depend on $\lambda_b$. In other words, $I_1$ and $D_1+I_1$ scaled by $\|X_1\|^{\alpha}$ do not depend on $\lambda_b$ so that densely deploying BSs does not influence the statistical performances of activating devices in this case. This observation motivates us to propose an energy-efficient power control for the devices and the BSs in the following to ameliorate the activation performance of each BS and the uplink coverage performance of each device. 

\section{Analysis of Total Activation Probability and Uplink Coverage}\label{Sec:SuccActProb}
As shown in the previous section, the total activation probability consists of the true activation probability and the false activation probability. In this section, we will first analyze the false, true and total activation probabilities and then study how they influence the uplink coverage performance of a device that is correctly activated. Recall that the false activation probability represents how likely the IoT devices are accidentally activated by the interfering active BSs. Thus, the devices that are accidentally activated will interfere the uplink transmissions of the devices that are correctly activated. In the other words, the false activation probability considerably affects the uplink transmission performance of the devices. We will finally provide numerical results to validate the derived analytical results pertaining to the total activation probability and the uplink coverage (probability) of a device.  

\subsection{Analysis of the Total Activation Probability} \label{SubSec:AnaActProb}
In this subsection, we study the false, true and total activation probabilities in a simple context where no any specific techniques for the devices and BSs (such as power control and BS coordination) are employed in the network. The false, true and total activation probabilities in this context are found in the following proposition.
\begin{proposition}\label{Prop:ActProb}
According to the total activation probability defined in \eqref{Eqn:DefActProb}, the false activation probability is found as
\begin{align}\label{Eqn:FalseActProb}
q_a = 1-\int_0^{\theta_a}   \mathcal{L}^{-1}\left\{\int_0^{\infty}\pi\lambda_b e^{-\pi\lambda_b[x+\mu\mathfrak{I}\left(x,s\overline{P}\right)]}\dif x\right\}(\tau)  \dif\tau,
\end{align}
where $\mathcal{L}^{-1}\{F(s)\}(\tau)$ denotes the inverse Laplace transform of function $F(s)$ with parameter $\tau>0$. The true activation probability is derived as
\begin{align}\label{Eqn:TrueActProb}
p_a = & 1- \frac{1}{\mu}\int_0^{\theta_a}\mathcal{L}^{-1}\left\{e^{-\pi\lambda_b\mu\mathfrak{I}(0,s\overline{P})}\right\}(\tau)\dif\tau \nonumber\\
&+\frac{(1-\mu)}{\mu} (1-q_a).
\end{align}
Thus, the total activation probability defined in \eqref{Eqn:DefActProb} is given by
\begin{align}
\eta_a = 1-\int_0^{\theta_a} \mathcal{L}^{-1}\left\{e^{-\pi\lambda_b\mu\mathfrak{I}(0,s\overline{P})}\right\}(\tau)\dif\tau. \label{Eqn:ActProb}
\end{align}
\end{proposition}
\begin{IEEEproof}
See Appendix \ref{App:ProofActProb}.
\end{IEEEproof}
\noindent According to the identity of $\mathfrak{I}(x,y)$ in \eqref{Eqn:IdenJfun} and the approximate result of $\mathcal{L}_{I_1}(s)$ in \eqref{Eqn:LapTransIntFirst}, the false activation probability in \eqref{Eqn:FalseActProb} can be approximated by
\begin{align}\label{Eqn:ApproxFalseActProb}
q_a \approx 1-\frac{1}{1-\mu\epsilon_1}\int_0^{\theta_a}\mathcal{L}^{-1}\left\{e^{-\pi\mu\lambda_b\mathfrak{I}(0,s\overline{P})}\right\}(\tau) \dif\tau,
\end{align}
whereas the true activation probability in \eqref{Eqn:TrueActProb} is approximately given by
\begin{align}\label{Eqn:ApproxTrueActProb}
p_a \approx 1-\frac{1-\epsilon_1}{1-\mu\epsilon_1}\int_0^{\theta_a}\mathcal{L}^{-1}\left\{e^{-\pi\lambda_b\mu\mathfrak{I}(0,s\overline{P})}\right\}(\tau)\dif\tau.
\end{align}
For the special case of $\alpha=4$, the integral in \eqref{Eqn:ApproxFalseActProb} and \eqref{Eqn:ApproxTrueActProb} has a closed-form result given by \cite{MAIA12}
\begin{align*}
&\int_0^{\theta_a}\mathcal{L}^{-1}\left\{e^{-\pi\mu\lambda_b\mathfrak{I}(0,s\overline{P})}\right\}(\tau) \dif\tau\nonumber\\ 
&= \mathrm{erfc}\left(\frac{\pi^{\frac{3}{2}}\lambda_b\mu \Gamma(m+\frac{1}{2})\sqrt{\overline{P}}}{2\sqrt{\theta_a}\Gamma(m)}\right),
\end{align*}
where $\mathrm{erfc}(z)\defn \frac{2}{\sqrt{\pi}}\int_z^{\infty} e^{-t^2}\dif t$ is the complementary error function with argument $z\geq 0$. In light of this, when $\alpha=4$, $q_a$  in \eqref{Eqn:ApproxFalseActProb} and $p_a$ in \eqref{Eqn:ApproxTrueActProb}  reduce to the following approximated closed-form results
\begin{align}
q_a
\approx 1-\frac{1}{(1-\epsilon_1\mu)}\mathrm{erfc}\left(\frac{\pi^{\frac{3}{2}}\lambda_b\mu \Gamma(m+\frac{1}{2})\sqrt{\overline{P}}}{2\sqrt{\theta_a}\Gamma(m)}\right)\label{Eqn:FalseActProbAlpha4}
\end{align} 
and 
\begin{align}
p_a 
\approx 1-\frac{(1-\epsilon_1)}{1-\epsilon_1\mu}\mathrm{erfc}\left(\frac{\pi^{\frac{3}{2}}\lambda_b\mu \Gamma(m+\frac{1}{2})\sqrt{\overline{P}}}{2\sqrt{\theta_a}\Gamma(m)}\right), \label{Eqn:TrueActProbAlpha4}
\end{align} 
respectively. $\eta_a$ in \eqref{Eqn:ActProb} reduces to the following exact closed-form expression:
\begin{align}
\eta_a =\mathrm{erf}\left(\frac{\pi^{\frac{3}{2}}\lambda_b\mu \Gamma(m+\frac{1}{2})\sqrt{\overline{P}}}{2\sqrt{\theta_a}\Gamma(m)}\right), \label{Eqn:ActProbAlpha4}
\end{align}
where $\mathrm{erf}(z)=1-\mathrm{erfc}(z)$. To the best of our knowledge, the nearly closed-form results in \eqref{Eqn:FalseActProbAlpha4}-\eqref{Eqn:ActProbAlpha4} that work for different fading models are first derived in this paper. Although the closed-form expressions of $q_a$, $p_a$ and $\eta_a$ cannot be found for the case of $\alpha\neq 4$, we can still evaluate them by resorting to numerical techniques. 

From the above results of $q_a$, $p_a$ and $\eta_a$, we clearly learn how network parameters $\lambda_b$, $\theta_a$, $\mu$, $\epsilon_1$, and $\overline{P}$ impact the statistics of the activation process. It is fairly easy to see that the three probabilities $p_a$, $q_a$ and $\eta_a$ are all dominated by the term $\lambda_b\mu(\overline{P}/\theta_a)^{\frac{2}{\alpha}}$. This indicates that increasing $\lambda_b$, $\mu$ and $\overline{P}$ increases $p_a$, $q_a$ and $\eta_a$ whereas increasing $\theta_a$ reduces them. In other words, the methods of densely deploying BSs, frequently activating the devices and using large transmit power all improve these three probabilities. However, none of these methods is guaranteed to ameliorate the activation performance, i.e., simultaneously decreasing the false activation probability and increasing the true activation probability. A possible approach to attaining a high activation performance is to reduce $\epsilon_1$  because \eqref{Eqn:FalseActProbAlpha4} and \eqref{Eqn:TrueActProbAlpha4} indicate that $q_a$ monotonically decreases and $p_a$ monotonically increases as $\epsilon_1$ increases. Furthermore, from the proof of Proposition \ref{Prop:LapWakeUpSignal} and Remark \ref{Rem:SpeFunInterference}, we realize that $\epsilon_1$ is pertaining to the term $D_1$ in \eqref{Eqn:DefActProb} and we can thus increase $\epsilon_1$ by augmenting $D_1$. Accordingly, a promising method to augment $D_1$ in the downlink is to perform \textit{downlink power control} and/or \textit{BS coordination}. Such a method and its related statistics of the activation signal process will be studied in Section \ref{SubSec:ActProbAdaActThreshold}. Next, we will analyze the uplink coverage (probability) of the activated devices and see how it is affected by $p_a$, $q_a$ and $\eta_a$. Afterwards, we will provide numerical results in Section \ref{SubSec:NumSuccActProb} to validate the above analytical results of $p_a$, $q_a$ and $\eta_a$.

\subsection{Analysis of the Uplink Coverage}\label{SubSec:ActProbAdaActThreshold}
Suppose each of the devices in a cell is allocated a unique resource block (RB) for uplink transmission and each of them begins uplink transmission once activated. Now consider the typical device is activated and then begin to transmit data to BS $X_1$. Consider the cellular network is dense and interference-limited so that the signal-to-interference (SIR) at BS $X_1$ can be written as\footnote{In practice, a densely-deployed cellular network is interference-limited so that the thermal noise power is very small if compared with the interference in the network. Moreover, using SIR to analyze the SIR-based performance metrics is much more tractable than using signal-to-interference-plus noise ratio (SINR) in a Poisson cellular network, which was pointed out in the literature \cite{CHL19}. In light of the above two reasons, we decide to adopt SIR instead of SINR to analyze the uplink coverage in the following.}
\begin{align}\label{Eqn:DefUplinkSIR}
\gamma_1 = \frac{Q_1H_1\|X_1\|^{-\alpha}}{\sum_{j:D_j\in\Phi_{d,a}}Q_jH_j\|D_j-X_1\|^{-\alpha}},
\end{align}  
where $Q_j$ is the transmit power of device $D_j$ and $\Phi_{d,a}\subseteq\Phi_d$ denotes the set of the activated devices using the same RB as the typical device. All $Q_j$'s are i.i.d. if they are random. Since the activation events of all the devices in the network are independent, all the activated devices form a thinning homogeneous PPP with density $\eta_a\lambda_d$. According to \eqref{Eqn:DisNumDev}, the PMF of the number of the activated devices in a cell can be inferred as
\begin{align}
\phi_{n,a} =\frac{\Gamma(n+\frac{7}{2})}{n!\Gamma(\frac{7}{2})}\left(1+\frac{2\eta_a\lambda_d}{7\lambda_b}\right)^{-(n+\frac{7}{2})},
\end{align}
and the probability that there are no devices activated in a cell is $\phi_{0,a} = (1+7\eta_a\lambda_d/2\lambda_b)^{-\frac{7}{2}}$ so that the density of the BSs with at least one activated device is $(1-\phi_{0,a})\lambda_b$ and the average number of the activated devices in a cell is $\eta_a\lambda_d/\lambda_b$. Note that each cell almost surely has at least one activated device provided $\eta_a\lambda_d/\lambda_b$ is very large (i.e., $\phi_{0,a}\approx 0$). Suppose the probability that a device is allocated to use any one of the RBs in a cell is the same and equal to $\rho\in(0,1)$. Hence, the density of $\Phi_{d,a}$ is $\rho\eta_a\lambda_d$ in that all the activated devices in the same cell use different RBs to transmit their data. 

According to \eqref{Eqn:DefUplinkSIR}, the uplink coverage (probability) of an activated device is defined as
\begin{align}
\eta_c \defn \mathbb{P}\left[\gamma_1\geq \theta_c\right],
\end{align}
where $\theta_c>0$ is the SIR threshold for successful decoding. It is explicitly derived as shown in the following proposition. 
\begin{proposition}\label{Prop:UplinkCoveProb}
If $m$ is a non-negative integer and the fractional moment $\mathbb{E}[Q^a]$ of $Q_j$ exists for $a\in(0,1)$ and all $j\in\mathbb{N}_+$, then the uplink coverage of the activated devices can be found as
\begin{align}\label{Eqn:UplinkCoveProb}
\eta_c = \frac{\dif^{m-1}}{\dif\tau^{m-1}}\mathbb{E}\bigg\{&\frac{\tau^{m-1}}{(m-1)!}\exp\bigg[-\pi\rho\eta_a\lambda_d\mathfrak{I}(0,1)\mathbb{E}\left[Q^{\frac{2}{\alpha}}\right]\nonumber\\
&\|X_1\|^2\left(\frac{m}{\tau Q}\right)^{\frac{2}{\alpha}}\bigg]\bigg\}\bigg|_{\tau=\theta^{-1}_c},
\end{align}
where $\mathbb{E}\{\cdot\}$ is the operator for taking the average of $\|X_1\|^2$ and $Q^{-\frac{2}{\alpha}}$. 
\end{proposition}
\begin{IEEEproof}
See Appendix \ref{App:ProofUplinkCoveProb}.
\end{IEEEproof}
\begin{remark}\label{Rem:UplinkCoverage}
For Rayleigh fading, $\eta_c$ in \eqref{Eqn:UplinkCoveProb} for $m=1$ reduces to
\begin{align}\label{Eqn:UplinkCoveProbRayFad}
\eta_c =\mathbb{E}\left\{\exp\left[-\pi\rho\eta_a\lambda_d\mathfrak{I}(0,1)\mathbb{E}\left[Q^{\frac{2}{\alpha}}\right]\|X_1\|^2\left(\frac{\theta_c}{Q}\right)^{\frac{2}{\alpha}}\right]\right\},
\end{align}
which is lower bounded as
\begin{align}
\eta_c \geq \exp\left\{-\pi\rho\eta_a\lambda_d\mathfrak{I}(0,1)\theta_c^{\frac{2}{\alpha}}\mathbb{E}\left[Q^{\frac{2}{\alpha}}\right]\mathbb{E}\left[\|X_1\|^2Q^{-\frac{2}{\alpha}}\right]\right\}
\end{align}
based on Jensen's inequality. Hence, if $Q$ is random and  $\mathbb{E}\left[Q^{\frac{2}{\alpha}}\right]\mathbb{E}\left[\|X_1\|^2Q^{-\frac{2}{\alpha}}\right]\leq \mathbb{E}\left[\|X_1\|^2\right]= \frac{1}{\pi\lambda_b}$ holds, the uplink coverage of the activated devices using stochastic transmit power is higher than that of the activated devices using constant transmit power. 
\end{remark}
Although the uplink coverage in Proposition \ref{Prop:UplinkCoveProb} is somewhat complex, it is very general since it considers stochastic transmit power used by all the activated devices. As pointed out in Remark~\ref{Rem:UplinkCoverage}, using stochastic transmit power for each device is able to enhance the uplink coverage. In addition, it is able to save transmit power, which can be explained as follows. First consider the context in which all the activated devices using the same constant transmit power $\overline{Q}$. In this context of constant uplink power control (i.e., no uplink power control), $\eta_c$ for Rayleigh fading  in \eqref{Eqn:UplinkCoveProbRayFad} reduces to the following closed-form result:
\begin{align}\label{Eqn:CovProbWithoutPowerControl}
\eta_c = \frac{\lambda_b}{\lambda_b+\rho\eta_a\lambda_d\theta_c^{\frac{2}{\alpha}}\mathfrak{I}(0,1)},
\end{align}
which is dominated by $\rho\eta_a\lambda_d\theta^{\frac{2}{\alpha}}_c/\lambda_b$. Now consider another context in which we propose the following uplink (energy-efficient) power control for the typical device
\begin{align}\label{Eqn:UplinkStocPowCon}
Q_1= \frac{\overline{Q}\|X_1\|^{\alpha\nu}}{\Gamma(1+\frac{\alpha\nu}{2})}&\bigg[(\pi\lambda^{\min}_b)^{\frac{\alpha\nu}{2}}\mathds{1}(\nu\geq 0)+(\pi\lambda^{\max}_b)^{\frac{\alpha\nu}{2}}\nonumber\\
&\times\mathds{1}\left(-\frac{2}{\alpha}<\nu< 0\right)\bigg],
\end{align}
where $\nu$ is a constant greater than $-\frac{2}{\alpha}$ so as to make $\mathbb{E}\left[Q^{\frac{2}{\alpha}}_1\right]$ and $\Gamma(1+\frac{\alpha\nu}{2})$ both exist, $\lambda^{\max}_b$ is the maximum (minimum) density for deploying the BSs, $\lambda^{\min}_b$ is the minimum density for deploying the BSs, and $\mathds{1}(\mathcal{A})$ is the indicator function that is unity if event $\mathcal{A}$ is true and zero otherwise. Note that $\lambda^{\min}_b\leq \lambda_b\leq \lambda^{\max}_b$. The mean of $Q_1$ is given by
\begin{align}\label{Eqn:AvgUplinkTranPower}
\mathbb{E}[Q] = \overline{Q}&\bigg[\left(\frac{\lambda^{\min}_b}{\lambda_b}\right)^{\frac{\alpha\nu}{2}}\mathds{1}(\nu\geq 0)+\left(\frac{\lambda_b}{\lambda^{\max}_b}\right)^{-\frac{\nu\alpha}{2}}\nonumber\\
&\times\mathds{1}\left(0>\nu>-\frac{2}{\alpha}\right)\bigg].
\end{align}
Note that the power control in \eqref{Eqn:UplinkStocPowCon} is energy-efficient if compared with constant (no) power control because it leads to $\mathbb{E}[Q]\leq \overline{Q}$ if  $\nu>-\frac{2}{\alpha}$. Using the proposed uplink power control, $\eta_c$ in \eqref{Eqn:UplinkCoveProb} for Rayleigh fading can be found as\footnote{Note that the uplink coverage in \eqref{Eqn:UplinkCovProbPowerCon} for $\nu=0$ (i.e., constant (no) power control) is similar to the uplink coverage in some prior works (e.g., see \cite{NKZDJGA18}). Thus, we can use the uplink coverage with $\nu=0$ as a baseline result to compare with the uplink coverage with $\nu\neq 0$.}
\begin{align}\label{Eqn:UplinkCovProbPowerCon}
\eta_c (\nu,n_a)
=&\int_0^{\infty} e^{-[n_a\Gamma(1+\nu)u^{(1-\nu)}+u]} \dif u,
\end{align}
where $n_a\defn \frac{\lambda_d}{\lambda_b}\rho\eta_a\theta^{\frac{2}{\alpha}}_c\mathfrak{I}(0,1)$. Note that $\eta_c(0,n_a)=1/(1+n_a)$ is equal to the uplink coverage in \eqref{Eqn:CovProbWithoutPowerControl} for constant (no) power control whereas $\eta_c(1,n_a)=\exp(-n_a)$ is the uplink coverage with \textit{channel inversion power control}. Since $e^{-n_a}< \frac{1}{1+n_a}$ for $n_a>0$, the uplink coverage performance of the channel inversion power control is essentially worse than that of the constant power control. Hence, we can conclude the following set of $(\nu,n_a)$
\begin{align}\label{Eqn:SetPowerCon}
\mathcal{V}_a\defn\bigg\{& (\nu,n_a)\in\mathbb{R}\times\mathbb{R}_{++} : \nu>-\frac{2}{\alpha},\nonumber\\
 &\eta_c(0,n_a)\leq \eta_c(\nu,n_a) \bigg\},
\end{align}
which allows a device using \eqref{Eqn:UplinkStocPowCon} to achieve higher uplink coverage and consume lower average power for the Rayleigh fading case. In other words, when the set $\mathcal{V}_a$ is not empty, the activated devices that control their uplink transmit power by using \eqref{Eqn:UplinkStocPowCon} with $(\nu,n_a)\in\mathcal{V}_a$ can save more power and achieve higher uplink coverage than they merely use constant transmit power. We will numerically validate the above observations regarding the uplink coverage with the uplink energy-efficient power control in the following subsection.

\subsection{Numerical Results and Discussions}\label{SubSec:NumSuccActProb}

\begin{table*}[!t]
	\centering
	\caption{Network Parameters for Simulation}\label{Tab:SimPara}
	\begin{tabular}{|c|c|c|}
		\hline Parameter $\setminus$ Transmitter Type & Base Station (Downlink) & IoT Device (Uplink)\\ 
		\hline Transmit Power (W) $\overline{P}$, $ \overline{Q}$  & 20  & 0.2  \\ 
		\hline Density (BSs/m$^2$, devices/m$^2$) $\lambda_b$, $\lambda_d$  & $1\times 10^{-5}\sim 1.5\times 10^{-4}$ & $100\lambda_b\sim 500\lambda_b$ (or see figures)    \\ 
		\hline Activation Threshold ($\mu$W) $\theta_a$   & $0.1$ & not applicable    \\
		\hline SIR Threshold $\theta_c$   & not applicable &  $1$    \\
		\hline Active Probability of a BS $\mu$   & 0.25 &  not applicable    \\
		\hline Probability of Using Each RB $\rho$   & not applicable & $0.01$    \\
		\hline Path-loss Exponent $\alpha$ &\multicolumn{2}{c|}{4} \\ 
		\hline Minimum BS density (BSs/m$^2$) $\lambda^{\min}_b$ &\multicolumn{2}{c|}{$5\times 10^{-6}$} \\ 
		\hline Maximum BS density (BSs/m$^2$) $\lambda^{\max}_b$ &\multicolumn{2}{c|}{$3\times 10^{-4}$} \\
		\hline Nakagami-$m$ Fading &\multicolumn{2}{c|}{ $m=1$ (Rayleigh Fading)} \\
		\hline 
	\end{tabular} 
\end{table*}

\begin{figure*}[!t]
	\centering
	\includegraphics[width=\textwidth,height=2.8in]{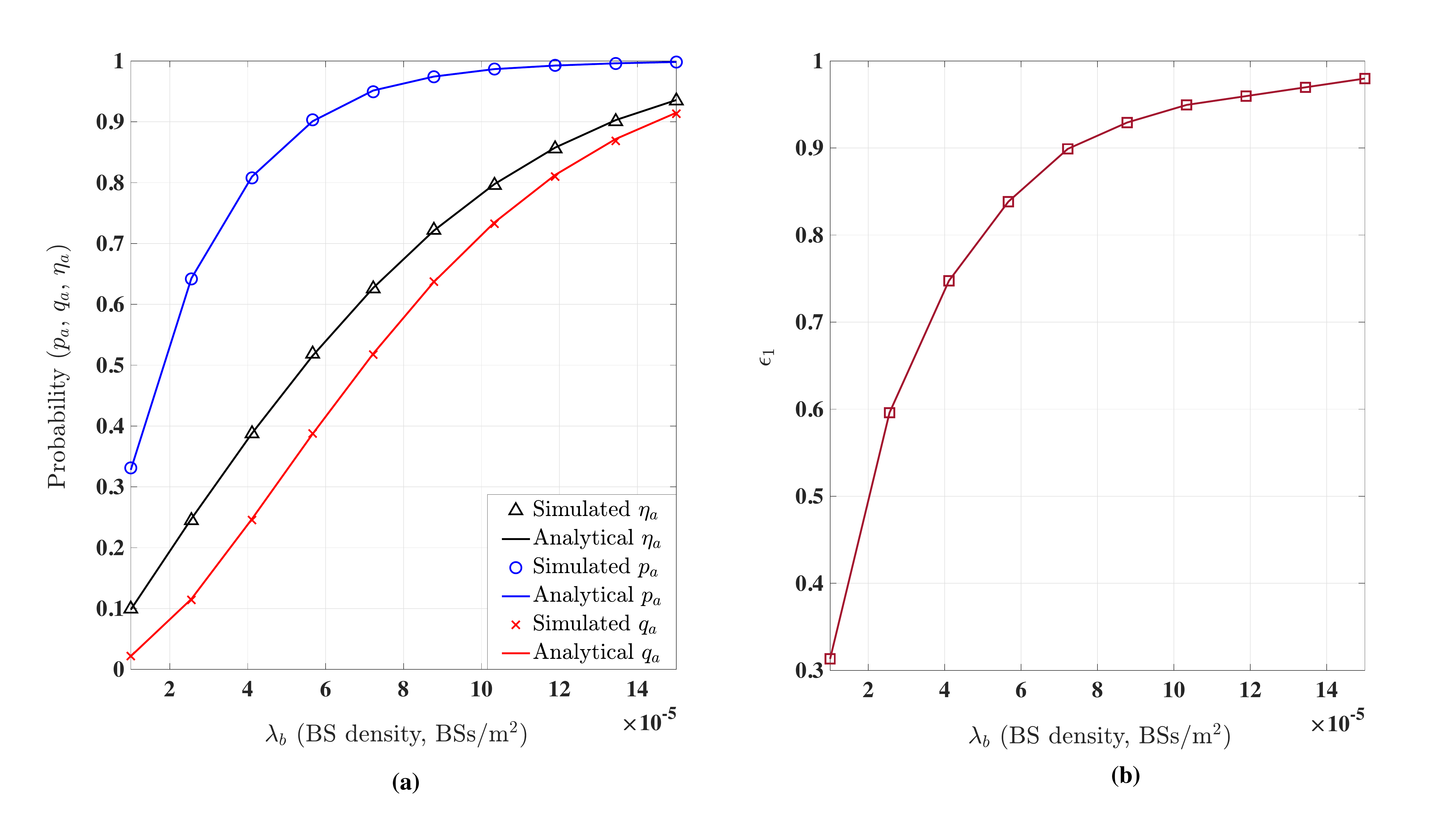}
	\caption{Simulation results of the activation process: (a) activation probability $\eta$, true activation probability $p_a$ and false activation probability $q_a$ versus $\lambda_b$, (b) $\epsilon_1$ versus $\lambda_b$.}
	\label{Fig:ActiveProb}
\end{figure*}

In this subsection, simulation results are provided to validate the analyses and observations of the activation probabilities and the uplink coverage in the previous subsections. The network parameters used for simulation are shown in Table \ref{Tab:SimPara}. We first show the simulation results of the activation process in Fig. \ref{Fig:ActiveProb}. In Fig. \ref{Fig:ActiveProb}(a), we can see all the simulated results of the false activation probability and the true activation probability coincide with their corresponding analytical results in \eqref{Eqn:FalseActProbAlpha4} and \eqref{Eqn:TrueActProbAlpha4} that are found by using the values of $\epsilon_1$ for different BS densities in Fig. \ref{Fig:ActiveProb}(b).  Thus, \eqref{Eqn:FalseActProbAlpha4} and \eqref{Eqn:TrueActProbAlpha4} are very accurate. In addition, \eqref{Eqn:ActProbAlpha4} is correct since the simulation results of the activation probability perfectly coincide with the analytical results calculated by it. As can be seen in Fig. \ref{Fig:ActiveProb}(a), the true activation probability $p_a$ is much higher than the false activation probability $q_a$ for all the BS densities, which is good for the network since a device is more likely to be activated correctly. However, $q_a$ grows much faster than $p_a$ as the BS density increases. For example, $p_a=0.75$ and $q_a=0.2$ when $\lambda_b=8\times 10^{-5}$ BSs/m$^2$ whereas $p_a=0.89$ and $q_a=0.36$ when $\lambda_b$ increases up to $1.2\times 10^{-4}$ BSs/m$^2$. In this case, $q_a$ increases by $80\%$, while $p_a$ merely grows by $18.7\%$, which means the activation performance is getting worse as the network is getting denser. The reason that the activation performance becomes worse as the network gets denser is because the aggregated power of the interfering activation signals increases much faster than the power of the desired activation signal. In the following section, we will propose a downlink power control and BS coordination scheme to improve the activation performance.

\begin{figure*}[!t]
	\centering
	\includegraphics[width=\textwidth,height=2.8in]{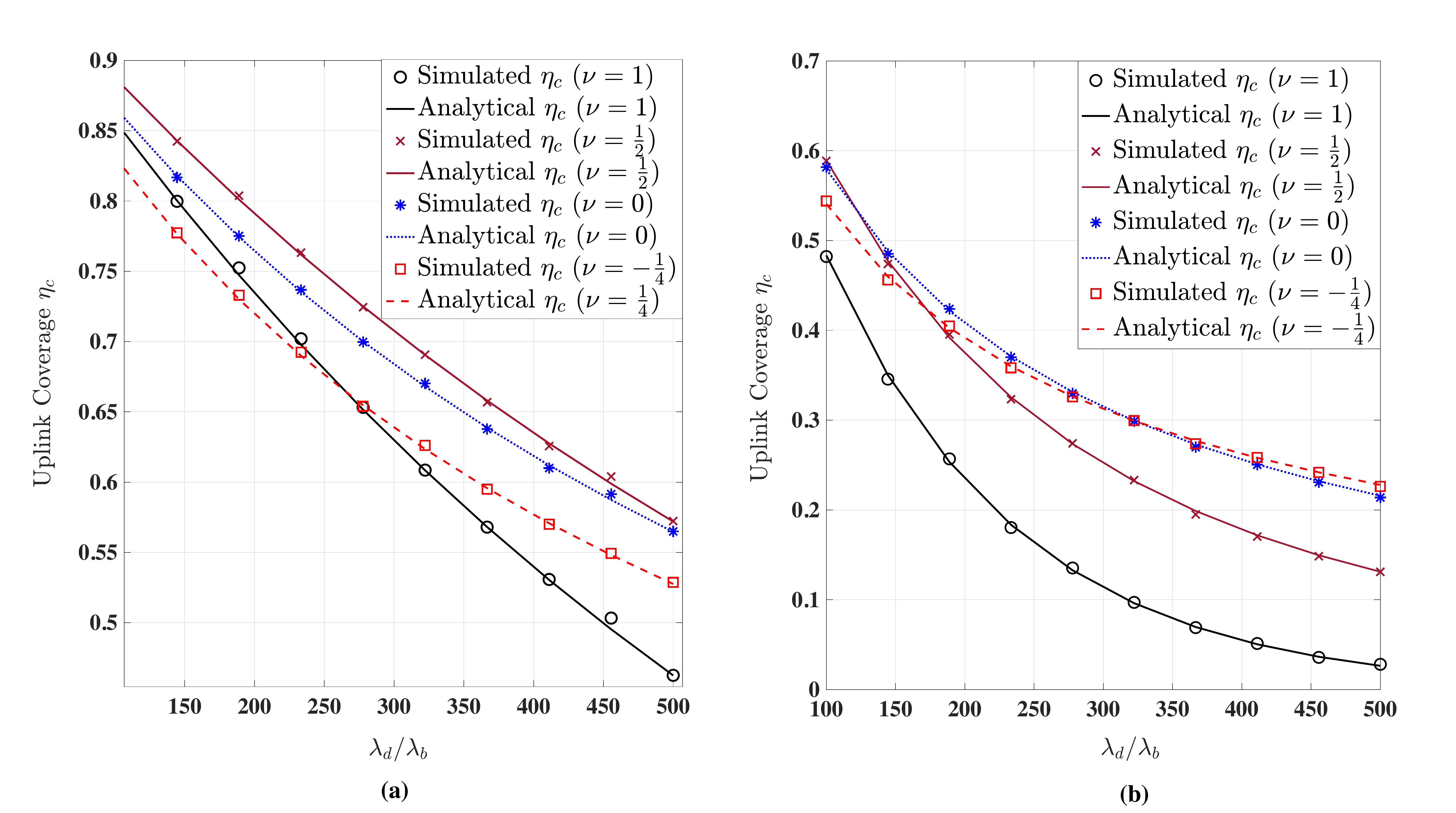}
	\caption{Simulation results of the uplink coverage probability with the proposed uplink power control for two different values of $\lambda_b$: (a) $\lambda_b=1\times 10^{-5}$ (BSs/m$^2$), (b) $\lambda_b=5\times 10^{-5}$ (BSs/m$^2$).}
	\label{Fig:UplinkCovProb}
\end{figure*}

\begin{figure*}[!t]
	\centering
	\includegraphics[width=\textwidth,height=2.8in]{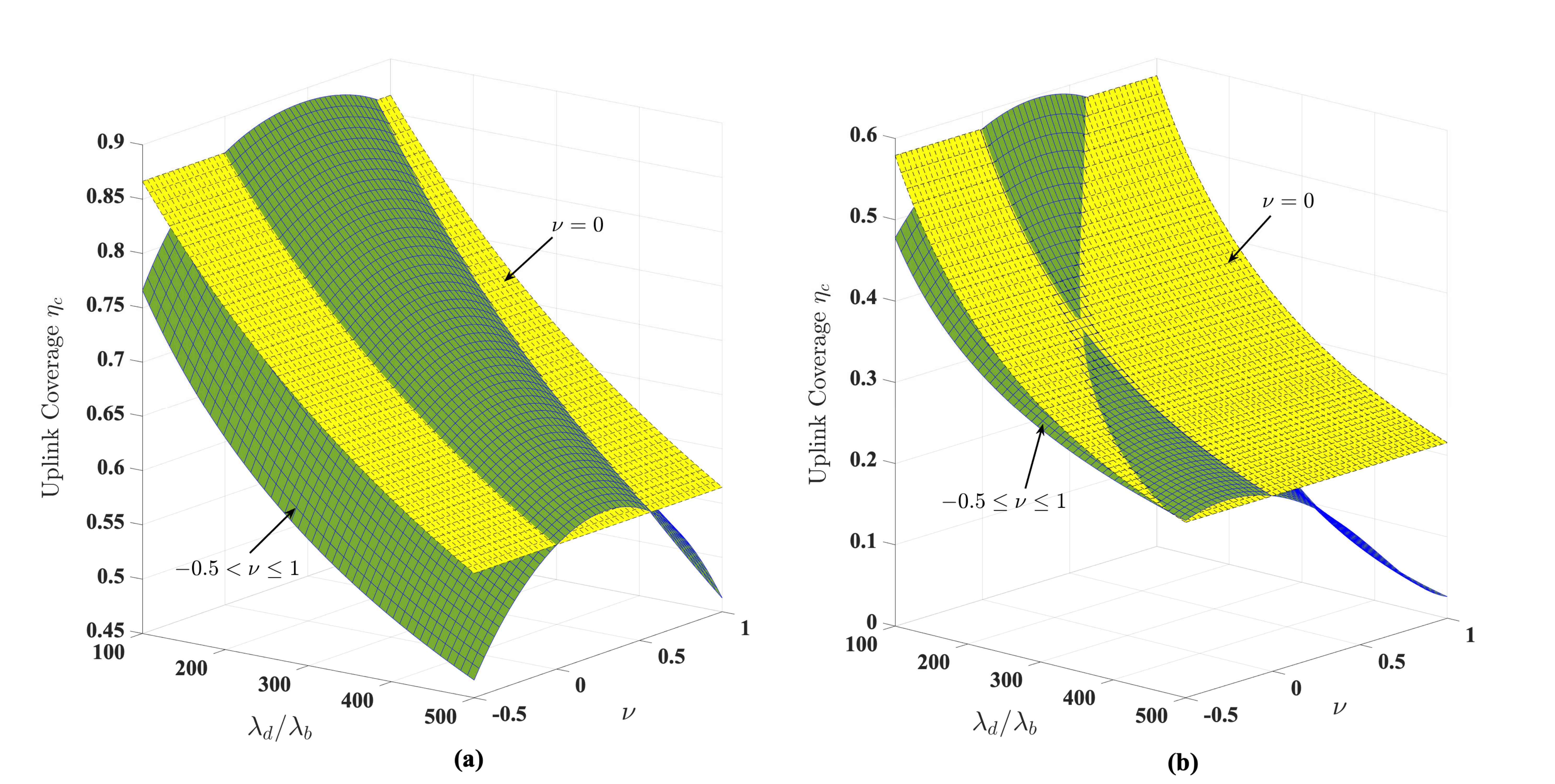}
	\caption{Simulation results of the uplink coverage probability with the proposed uplink power control for two different values of $\lambda_b$: (a) $\lambda_b=1\times 10^{-5}$ (BSs/m$^2$), (b) $\lambda_b=5\times 10^{-5}$ (BSs/m$^2$).}
	\label{Fig:UplinkCovprob3D}
\end{figure*}

Fig. \ref{Fig:UplinkCovProb} shows the simulation results of the uplink coverage using the uplink power control in~\eqref{Eqn:UplinkStocPowCon} for different values of $\frac{\lambda_d}{\lambda_b}$ which represents the average number of the devices associated with a BS. As we can see in the figure, all the analytical results are calculated by using \eqref{Eqn:UplinkCovProbPowerCon} and they perfectly coincide with their corresponding simulated results so that \eqref{Eqn:UplinkCovProbPowerCon} is correct. Also, we can observe that the constant uplink power control with $\nu=0$ does not always outperform the proposed uplink power control in terms of the uplink coverage. Changing $\nu$ significantly impacts the uplink coverage performance of the proposed uplink power control. In the case of $\lambda_b=1\times 10^{-5}$ BSs/m$^2$ shown in Fig.~\ref{Fig:UplinkCovProb}(a), for example, the uplink power control with $\nu=\frac{1}{2}$ is the only one that outperforms the constant power control in the simulation range of $\frac{\lambda_d}{\lambda_b}$. Another case of $\lambda_b=5\times 10^{-5}$ BSs/m$^2$ shown in Fig. \ref{Fig:UplinkCovProb}(b) further reveals that the proposed uplink power control is not always superior to the constant power control and we should properly choose the value of $\nu$ based on $\lambda_b$ as well as $\frac{\lambda_d}{\lambda_b}$. For instance, the proposed uplink power control with $\nu=0.5$ outperforms the constant power control when $100\leq \frac{\lambda_d}{\lambda_b}\leq 320$, yet it does not perform as well as the constant power control when $\frac{\lambda_d}{\lambda_b}>320$. Furthermore, we notice that the channel inversion power control (i.e., the case of $\nu=1$)  does not outperform the constant power control so that it is in general not a good option for uplink power control. To visually demonstrate whether there exist some combinations of $\nu$ and $\frac{\lambda_d}{\lambda_b}$ that allow the proposed power control to save power and enhance the uplink coverage, the three-dimensional simulation results that show how the uplink coverage varies with $\nu$ and $\frac{\lambda_d}{\lambda_b}$ are presented in Fig. \ref{Fig:UplinkCovprob3D}. As shown in Fig. \ref{Fig:UplinkCovprob3D}(a), for example, there indeed exists a set of $\nu$ and $\frac{\lambda_d}{\lambda_b}$ where the proposed uplink power control outperforms the constant power control so that it can reduce the power consumption and achieve a higher uplink coverage. In other words, we demonstrate that the set in \eqref{Eqn:SetPowerCon} is not empty and indeed exists.  

\section{Activation Probability and Uplink Coverage with Energy-Efficient Power Control and BS Coordination}\label{Sec:EngEffPowConBScoor}
The analytical results of the true activation and false activation probabilities found in Section~\ref{SubSec:AnaActProb} reveal that a possible approach to increasing the true activation probability and reducing the false activation probability at the same time is to adopt downlink power control and BS coordination. In this section, we will first propose a joint downlink (energy-efficient) power control and BS coordination scheme. We then  analyze the true and false activation probabilities and the uplink coverage under such a scheme and show that the proposed scheme is able to improve the activation and uplink coverage performances. Finally, some numerical results will be provided to validate our findings. 

\subsection{Activation Process with Downlink Power Control and BS Coordination}\label{SubSec:ActProbAdaBSCoor}

According to the uplink power control in \eqref{Eqn:UplinkStocPowCon}, it motivates us to propose the following downlink (energy-efficient) power control for BS $X_1$ associating with the typical device:
\begin{align}\label{Eqn:DownLinkPowCon}
P_1 \defn  \frac{\overline{P}\|X_1\|^{\alpha\beta}}{\Gamma(1+\frac{\alpha\beta}{2})}\bigg[& (\pi\lambda^{\min}_b)^{\frac{\alpha\beta}{2}}\mathds{1}(\beta\geq 0)+(\pi\lambda^{\max}_b)^{\frac{\alpha\beta}{2}}\nonumber\\
&\times\mathds{1}\left( -\frac{2}{\alpha}<\beta<0\right)\bigg],
\end{align} 
where $\beta$ is a constant greater than $-\frac{2}{\alpha}$ so as to let $\mathbb{E}[P^{\frac{2}{\alpha}}]$ and $\Gamma(1+\frac{\alpha\beta}{2})$ both exist. Again notice that $P_1=\overline{P}$ for $\beta=0$ is the case of constant (no) downlink power control. All the other BSs in the network also independently adopt the same power control scheme as BS $X_1$ so that all $P_i$'s are i.i.d. for all $i\in\mathbb{N}_+$. Likewise, the power control in \eqref{Eqn:DownLinkPowCon} is able to save power for a BS if compared with constant power control since $\mathbb{E}[P_1]=\overline{P}[(\frac{\lambda^{\min}_b}{\lambda_b})^{\frac{\alpha\beta}{2}}\mathds{1}(\beta\geq 0)+(\frac{\lambda^{\max}_b}{\lambda_b})^{\frac{\alpha\beta}{2}}\mathds{1}(-\frac{2}{\alpha}<\beta<0)]\leq \overline{P}$ so that it is energy-efficient. Moreover, it can reduce the impact from the BS density on the uplink coverage by properly choosing the value of exponent $\beta$ owing to $\|X_1\|^2\sim\exp(\pi\lambda_b)$. 

Suppose now the first $K$ nearest BSs to the typical device can be coordinated and synchronized to send their activation signals at the same time by using the downlink power control in \eqref{Eqn:DownLinkPowCon}. Such a joint downlink power control and BS coordination scheme results in  another activation signaling process at the typical device similar to $W$, termed the $K$th-coordinated activation signaling process, which is defined in the following:
\begin{align}\label{Eqn:CoordActProc}
W_{K} \defn \omega_1\underbrace{\left[\sum_{k=1}^{K}P_kH_k\|X_k\|^{-\alpha}\right]}_{D_K}+\underbrace{\sum_{i:X_i\in\Phi^K_b} \omega_iP_iH_i\|X_i\|^{-\alpha}}_{I_K},
\end{align}
where $X_k$ represents the $k$th nearest BS in $\Phi_b$ to the typical device and $\Phi_{b,K}\defn \Phi_b\setminus \{X_k\}^K_{k=1}$. Note that  we have $\omega_1=\cdots=\omega_K$ in \eqref{Eqn:CoordActProc} because the first $K$ nearest BSs to the typical device are coordinated to broadcast their activation signals at the same time. In light of this, $D_K$ represents the sum power of the desired activation signals from the first $K$ nearest BSs whereas $I_K$ stands for the aggregated power of the interfering signals from the BSs in $\Phi_{b,K}$. In other words, $W$ in \eqref{Eqn:ShotWakeUpPower} is essentially equal to $W_K$ in \eqref{Eqn:CoordActProc} for $K=1$. The Laplace transform of $W_K$, which will significantly facilitate our following analyses, can be found as shown in the following proposition.
\begin{proposition}\label{Prop:LapKthCoorActProc}
	For the $K$th-coordinated activation signaling process defined in \eqref{Eqn:CoordActProc}, if the fractional moments of $P_i$ and $H_i$ exist, then there exists an $\epsilon_K\in[0,1)$ such that the Laplace transform of $I_K$ can be approximated by
	\begin{align}\label{Eqn:LapTranI_K}
	\mathcal{L}_{I_K}(s) \approx\frac{\exp\left(-\pi\lambda_b\mu\mathfrak{I}(0,s)\mathbb{E}\left[P^{\frac{2}{\alpha}}\right]\right)}{\left(1-\mu\epsilon_K\right)^{K}},
	\end{align}
	where $\mathbb{E}\left[P^{\frac{2}{\alpha}}\right]$ is given by
	\begin{align}
	\mathbb{E}\left[P^{\frac{2}{\alpha}}\right]=&\left(\frac{\overline{P}[\Gamma\left(1+\beta\right)]^{\frac{\alpha}{2}}}{\Gamma(1+\frac{\alpha\beta}{2})}\right)^{\frac{2}{\alpha}} \bigg[ \left(\frac{\lambda^{\min}_b}{\lambda_b}\right)^{\beta}\mathds{1}(\beta\geq 0)\nonumber\\
	&+\left(\frac{\lambda^{\max}_b}{\lambda_b}\right)^{\beta}\mathds{1}\left( -\frac{2}{\alpha}<\beta<0\right)\bigg], \label{Eqn:FracMomDowlinkPower}
	\end{align}
	and the Laplace transform of $D_K+I_K$ is approximately derived as
	\begin{align}\label{Eqn:LapTranD_K}
	\mathcal{L}_{D_K+I_K}(s) \approx &[1-(1-\mu)\epsilon_K]^K\nonumber\\
	&\times\exp\left(-\pi\lambda_b\mu\mathfrak{I}(0,s)\mathbb{E}\left[P^{\frac{2}{\alpha}}\right]\right).
	\end{align}
 Using \eqref{Eqn:LapTranI_K} and \eqref{Eqn:LapTranD_K}, the Laplace transform of $W_K$ can be found as
	\begin{align}\label{Eqn:LapTransWcK}
	\mathcal{L}_{W_K}(s) \approx& \left[\mu(1-(1-\mu)\epsilon_K)^K+\frac{(1-\mu)}{(1-\mu\epsilon_K)^K}\right]\nonumber\\
	&\exp\left(-\pi\lambda_b\mu\mathfrak{I}(0,s)\mathbb{E}\left[P^{\frac{2}{\alpha}}\right]\right).
	\end{align}
\end{proposition}
\begin{IEEEproof}
	See Appendix \ref{App:LapKthCoorActProc}.
\end{IEEEproof}
\begin{remark}\label{Rem:EpsilonKandDowPowCon}
Note that $\{\epsilon_K\}$ is a monotonic decreasing sequence for all $K\in\mathbb{N}_+$, i.e.,  $\epsilon_1>\epsilon_2>\cdots>\epsilon_K>\epsilon_{K+1}>\cdots$, based on Remark \ref{Rem:SpeFunInterference} and we thus have $\lim_{K\rightarrow\infty} \epsilon_K=0$. Also, note that we can have $\mathbb{E}[P^{\frac{2}{\alpha}}]\leq \overline{P}^{\frac{2}{\alpha}}$ by choosing an appropriate $\beta$ such that  $\Gamma(1+\beta)<[\Gamma(1+\frac{\alpha\beta}{2})]^{\frac{2}{\alpha}}$ holds. This indicates that the downlink power control is also able to reduce the interference $I_K$.
\end{remark}
According to the definition of $I_K$ in \eqref{Eqn:CoordActProc}, we can infer that $\mathcal{L}_{I_K}(s)>\mathcal{L}_{I_{K-1}}(s)>\cdots>\mathcal{L}_{I_1}(s)$ so that we know $(1-\mu\epsilon_K)^K\leq (1-\mu\epsilon_1)$ for all $K\in\mathbb{N}_+$ by comparing \eqref{Eqn:LapTranI_K} with \eqref{Eqn:LapTransIntFirst} and $(1-\mu\epsilon_K)^K$ decreases as $K$ increases. As a result, $[1-(1-\mu)\epsilon_K]^K$ decreases as $K$ increases so that $\mathcal{L}_{D_K+I_K}(s)$ decreases (i.e., $D_K+I_K$ almost surely increases) as $K$ increases. These observations mean that a device gains more activation signal power when its serving BS tries to activate it and it receives less interfering signal power otherwise. Thus, the proposed downlink power control and BS coordination scheme is able to improve the activation performance, which can be explicitly validated in the following proposition in which the true and false activation probabilities are explicitly derived.
\begin{proposition}\label{Prop:ActProbDowlinkPowCon}
If the downlink power control in \eqref{Eqn:DownLinkPowCon} is adopted by all the BSs and the first $K$ nearest BSs to each device are coordinated to activate their devices at the same time, the false activation probability is approximated by
\begin{align}\label{Eqn:FalseActProbPowConBSCor}
q_a \approx & 1- \frac{1}{\left(1-\mu\epsilon_K\right)^{K}}\times\nonumber\\
&\int_0^{\theta_a}   \mathcal{L}^{-1}\left\{\exp\left(-\pi\lambda_b\mu\mathfrak{I}\left(0,s\right)\mathbb{E}\left[P^{\frac{2}{\alpha}}\right]\right)\right\}(\tau)  \dif\tau,
\end{align}
the true activation probability is accurately found as
\begin{align}\label{Eqn:TrueActProbPowConBSCor}
p_a \approx& 1-  [1-(1-\mu)\epsilon_K]^K\times \nonumber\\
&\int_0^{\theta_a}   \mathcal{L}^{-1}\left\{\exp\left(-\pi\lambda_b\mu\mathfrak{I}\left(0,s\right)\mathbb{E}\left[P^{\frac{2}{\alpha}}\right]\right)\right\}(\tau)  \dif\tau,
\end{align}
and the activation probability is accurately given by
\begin{align}\label{Eqn:ActProbPowConBSCor}
\eta_a \approx 1-& \left[\mu(1-(1-\mu)\epsilon_K)^K+\frac{(1-\mu)}{(1-\mu\epsilon_K)^K}\right]\times\nonumber\\
&\int_0^{\theta_a}   \mathcal{L}^{-1}\left\{\exp\left(-\pi\lambda_b\mu\mathfrak{I}\left(0,s\right)\mathbb{E}\left[P^{\frac{2}{\alpha}}\right]\right)\right\}(\tau)  \dif\tau.
\end{align}
\end{proposition}
\begin{IEEEproof}
The proof is omitted since it is similar to the proof of Proposition \ref{Prop:ActProb}.
\end{IEEEproof}
As shown in Proposition \ref{Prop:ActProbDowlinkPowCon}, if $K$ increases, then $(1-\mu\epsilon_K)^{K}$ decreases and $q_a$ in \eqref{Eqn:FalseActProbPowConBSCor} thus reduces. The false probability in \eqref{Eqn:FalseActProb} without downlink power control and BS coordination is larger than that in \eqref{Eqn:FalseActProbPowConBSCor} for a sufficiently large $K$. Similarly, if $p_a$ in \eqref{Eqn:TrueActProbPowConBSCor} is compared with the true probability in \eqref{Eqn:TrueActProb}  without downlink power control and BS coordination, then it is certainly larger for a sufficiently large $K$ because $[1-(1-\mu)\epsilon_K]^K$ reduces as $K$ increases. Therefore, the proposed downlink power control and BS coordination scheme can improve the activation performance even though it uses smaller average transmit power than the constant power control. In addition, for $\alpha=4$, the results in Proposition \ref{Prop:ActProbDowlinkPowCon} reduce to the following approximated closed-form results:
\begin{align}
q_a \approx 1-\frac{1}{(1-\mu\epsilon_K)^K}\mathrm{erfc}\left(\frac{\pi^{\frac{3}{2}}\lambda_b\mu \Gamma(m+\frac{1}{2})\mathbb{E}[\sqrt{P}]}{2\sqrt{\theta_a}\Gamma(m)}\right), \label{Eqn:FalseActProbPowConBSCorA4}
\end{align}
\begin{align}
p_a \approx 1-[1-(1-\mu)\epsilon_K]^K\mathrm{erfc}\left(\frac{\pi^{\frac{3}{2}}\lambda_b\mu \Gamma(m+\frac{1}{2})\mathbb{E}[\sqrt{P}]}{2\sqrt{\theta_a}\Gamma(m)}\right), \label{Eqn:TrueActProbPowConBSCorA4}
\end{align}
and
\begin{align}
\eta_a \approx 1-& \left[\mu(1-(1-\mu)\epsilon_K)^K+\frac{(1-\mu)}{(1-\mu\epsilon_K)^K}\right]\times \nonumber\\ & \mathrm{erfc}\left(\frac{\pi^{\frac{3}{2}}\lambda_b\mu \Gamma(m+\frac{1}{2})\mathbb{E}[\sqrt{P}]}{2\sqrt{\theta_a}\Gamma(m)}\right). \label{Eqn:ActProbPowConBSCorA4}
\end{align}
Note that \eqref{Eqn:FalseActProbPowConBSCorA4}, \eqref{Eqn:TrueActProbPowConBSCorA4} and  \eqref{Eqn:ActProbPowConBSCorA4} reduce to \eqref{Eqn:FalseActProbAlpha4}, \eqref{Eqn:TrueActProbAlpha4} and \eqref{Eqn:ActProbAlpha4} for $K=1$, respectively.  Next, we will discuss how the downlink power control and BS coordination scheme impacts the uplink coverage performance and how to properly choose the parameter $\beta$ in \eqref{Eqn:DownLinkPowCon} to improve the activation performance.  

\subsection{Analysis of the Uplink Coverage and the Activation Performance Index}\label{SubSec:ActProbAdaPowerControl}
The uplink coverage derived in \eqref{Eqn:UplinkCovProbPowerCon} clearly shows that it improves as $\eta_a$ reduces. To make $\eta_a$ reduce by using the proposed downlink power control and BS coordination scheme, we must have $\frac{\dif \eta_a}{\dif \epsilon_K}>0$ as $K$ increases so that the following inequality has to hold
\begin{align*}
(1-\mu)\frac{\dif q_a}{\dif \epsilon_K}+\mu\frac{\dif p_a}{\dif \epsilon_K}>0
\end{align*}
since $\eta_a = (1-\mu)q_a+\mu p_a$. By employing \eqref{Eqn:FalseActProbPowConBSCor} and \eqref{Eqn:TrueActProbPowConBSCor} in the above inequality, we get the following
\begin{align}\label{Eqn:TotalActProbEpsilonKIneq}
 \left[1-(1-\mu)\epsilon_K\right]^{K-1}(1-\mu\epsilon_K)^{K+1}< 1,
\end{align}
which always holds for all $K\geq 1$ because $[1-(1-\mu)\epsilon_K]^{K-1}<1$ and $(1-\mu\epsilon_K)<1$. Therefore, $\frac{\dif \eta_a}{\dif \epsilon_K}>0$ for all $K\geq 1$ and we thus conclude that the proposed downlink power control and BS coordination scheme can benefit the uplink coverage. 

\begin{figure*}[!t]
	\centering
	\includegraphics[width=\textwidth,height=2.8in]{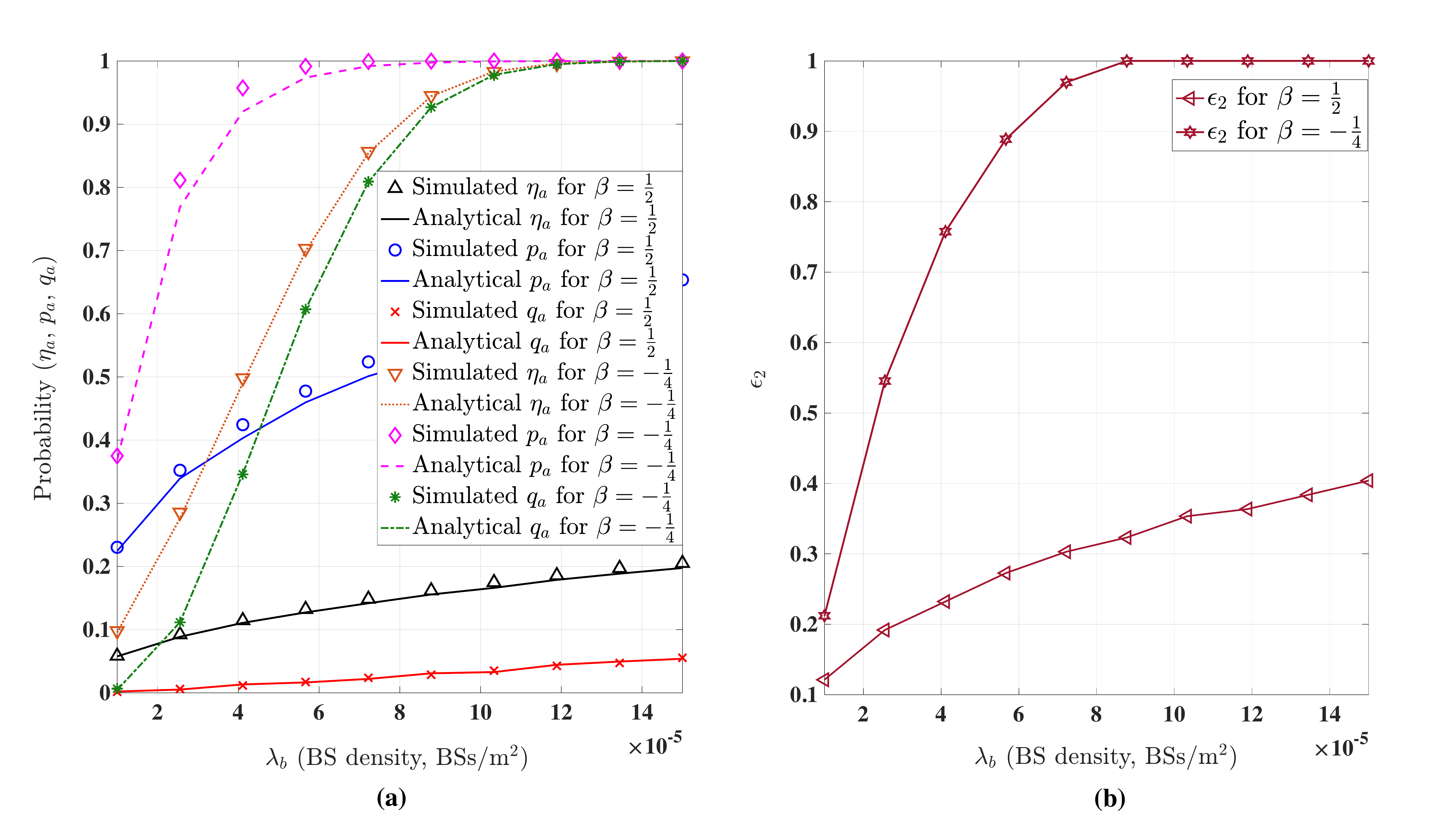}
	\caption{Simulation results of the second-coordinated activation process with the proposed downlink power control with different values of $\beta$: (a) Probabilities $\eta_a$, $p_a$, $q_a$ versus $\lambda_b$, (b) $\epsilon_2$ versus $\lambda_b$.}
	\label{Fig:ActiveProbPowCon_2BSCoor}
\end{figure*}

\begin{figure*}[!t]
	\centering
	\includegraphics[width=\textwidth,height=2.8in]{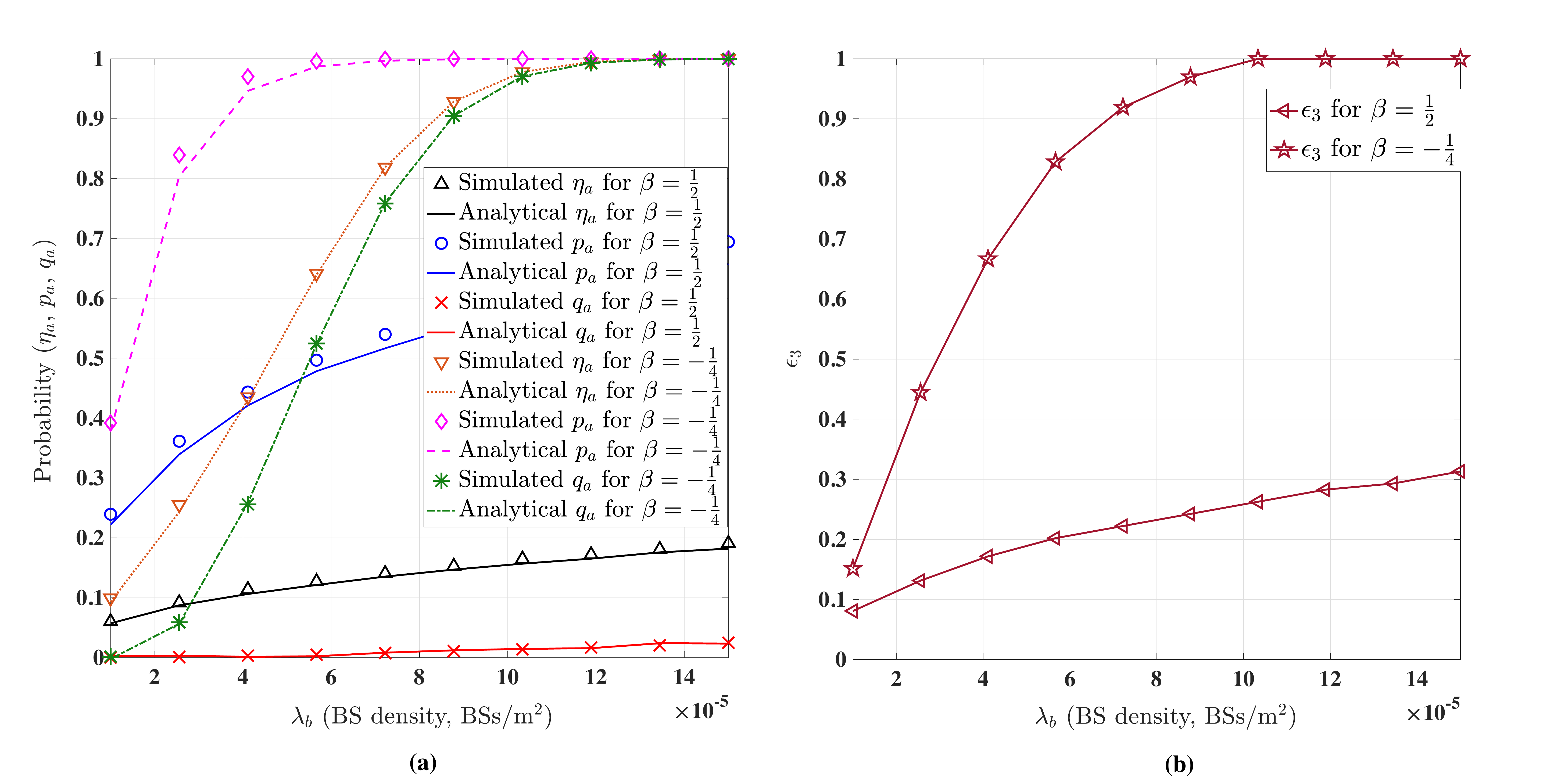}
	\caption{Simulation results of the third-coordinated activation process with the proposed downlink power control with different values of $\beta$: (a) Probabilities $\eta_a$, $p_a$, $q_a$ versus $\lambda_b$, (b) $\epsilon_3$ versus $\lambda_b$.}
	\label{Fig:ActiveProbPowCon_3BSCoor}
\end{figure*}

We hitherto have shown that the proposed downlink power control and BS coordination scheme not only improve the activation performance but also benefit the uplink coverage. To quantitatively evaluate the activation performance of the proposed downlink power control and BS coordination scheme, we define the activation performance index of an active BS as
\begin{align}\label{Eqn:ActPerIndex}
\zeta_a \defn (1-\eta_a)\left(\frac{p_a}{q_a}-1\right),
\end{align}  
which can be calculated by using \eqref{Eqn:FalseActProbPowConBSCor}, \eqref{Eqn:TrueActProbPowConBSCor} and \eqref{Eqn:ActProbPowConBSCor}. A large value of $\zeta_a$ reveals that an active BS has a high activation performance (i.e., it has large $p_a$ as well as small $q_a$ and $\eta_a$) because $1-\eta_a = 1-q_a[1+\mu(\frac{p_a}{q_a}-1)]$ reflects if $q_a$ and $\eta_a$ are small and the term $(\frac{p_a}{q_a}-1)$ reveals how much $p_a$ is larger than $q_a$. We can expound $\zeta_a$ in more detail as follows. Consider a case that each of the active BSs has $p_a$ which is much larger than $q_a$ that is not small. In this case, $\zeta_a$ is not large since $1-\eta_a$ is small and the active BSs thus do not have very good activation performance due to a large $q_a$ even though their $p_a$ is much larger than $q_a$. Moreover, $\zeta_a$ can also indicate whether the active BSs adopt too much downlink power since using large downlink power makes $\eta_a$, $p_a$ and $q_a$ approach to unity and may significantly lower $\zeta_a$. We will present numerical results in the following subsection to validate the above analysis and the aforementioned observations.

\subsection{Numerical Results and Discussions}
 
In this subsection, simulation results are provided to validate the previous analyses of the false and true activation probabilities, the uplink coverage and the activation performance index. The network parameters for simulation used in this section are the same as those in Table~\ref{Tab:SimPara}.  In Fig.~\ref{Fig:ActiveProbPowCon_2BSCoor}, we show the simulation results of the $K$th-coordinated activation signaling process $W_K$ for $K=2$ when the proposed downlink power control \eqref{Eqn:DownLinkPowCon} is employed. As can be seen in Fig.~ \ref{Fig:ActiveProbPowCon_2BSCoor}(a), the analytical results of $q_a$, $p_a$ and $\eta_a$ that are calculated by using \eqref{Eqn:FalseActProbPowConBSCorA4}, \eqref{Eqn:TrueActProbPowConBSCorA4} and \eqref{Eqn:ActProbPowConBSCorA4} with the values of $\epsilon_2$ for different BS densities in Fig. \ref{Fig:ActiveProbPowCon_2BSCoor}(b) are fairly close to their corresponding simulated results. Hence, the approximated results in \eqref{Eqn:FalseActProbPowConBSCorA4}, \eqref{Eqn:TrueActProbPowConBSCorA4} and \eqref{Eqn:ActProbPowConBSCorA4} are very accurate. Also, we observe that all $p_a$, $q_a$ and $\eta_a$ for the case of $\beta=-\frac{1}{4}$ are much larger than those for the case of $\beta=\frac{1}{2}$. This is because the average downlink power for $\beta=-\frac{1}{4}$ is larger than that for $\beta=\frac{1}{2}$. For the case of $K=3$, the simulation results of $q_a$, $p_a$ and $\eta_a$ are shown in Fig. \ref{Fig:ActiveProbPowCon_3BSCoor} and they can be compared with  Fig. \ref{Fig:ActiveProbPowCon_2BSCoor} to see some interesting differences between them. First, we can observe that $p_a$ increases and $q_a$ decreases as $K$ increases, as expected. Second, $\eta_a$ and $\epsilon_K$ both reduce as $K$ increases so that we validate that $\{\epsilon_K\}$ is a monotonic decreasing sequence (as pointed out in Remark \ref{Rem:EpsilonKandDowPowCon}) and $\frac{\dif\eta_a}{\dif \epsilon_K}>0$ (as shown in \eqref{Eqn:TotalActProbEpsilonKIneq}). Third, if we further compare Fig. \ref{Fig:ActiveProb} with Fig.~\ref{Fig:ActiveProbPowCon_2BSCoor} and Fig.~\ref{Fig:ActiveProbPowCon_3BSCoor}, we see that the activation performance is significantly improved by the proposed downlink power control and BS coordination scheme as $K$ increases from 1 to 2 whereas it slightly improved as $K$ increases from 2 to 3, which indicates that coordinating two or three BSs in practice would be good enough to significantly enhance the activation performance.

\begin{figure*}[!t]
	\centering
	\includegraphics[width=\textwidth,height=2.8in]{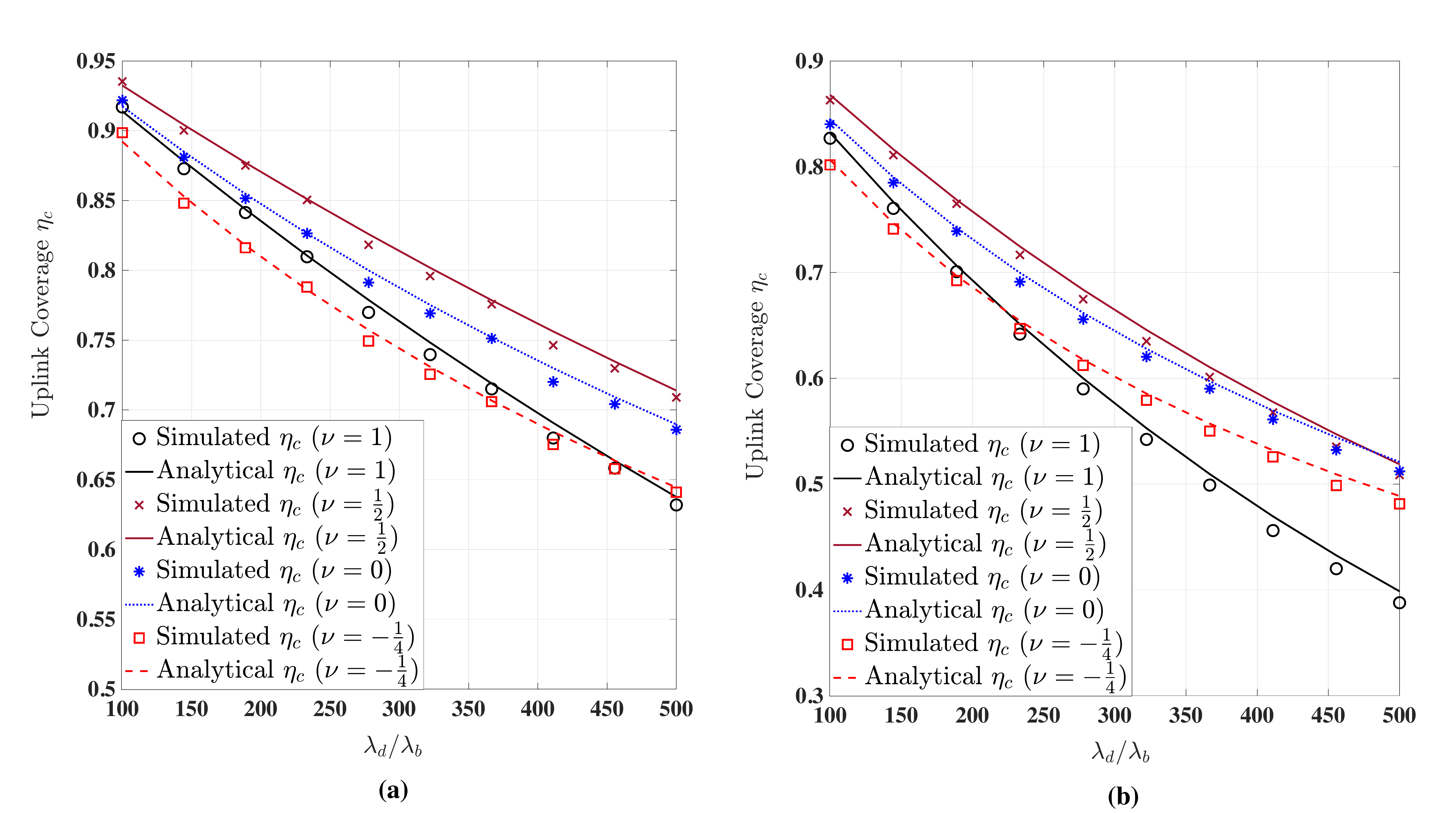}
	\caption{Simulation results of the uplink coverage probability with the proposed uplink power control and the proposed downlink power control for $\beta=\frac{1}{2}$ and BS coordination scheme with $K=3$: (a) the case of $\lambda_b=1\times 10^{-5}$ (BSs/m$^2$), (b) the case of $\lambda_b=5\times 10^{-5}$ (BSs/m$^2$).}
	\label{Fig:UplinkCovProbDowPowCon}
\end{figure*}

\begin{figure*}[!t]
	\centering
	\includegraphics[width=\textwidth,height=2.8in]{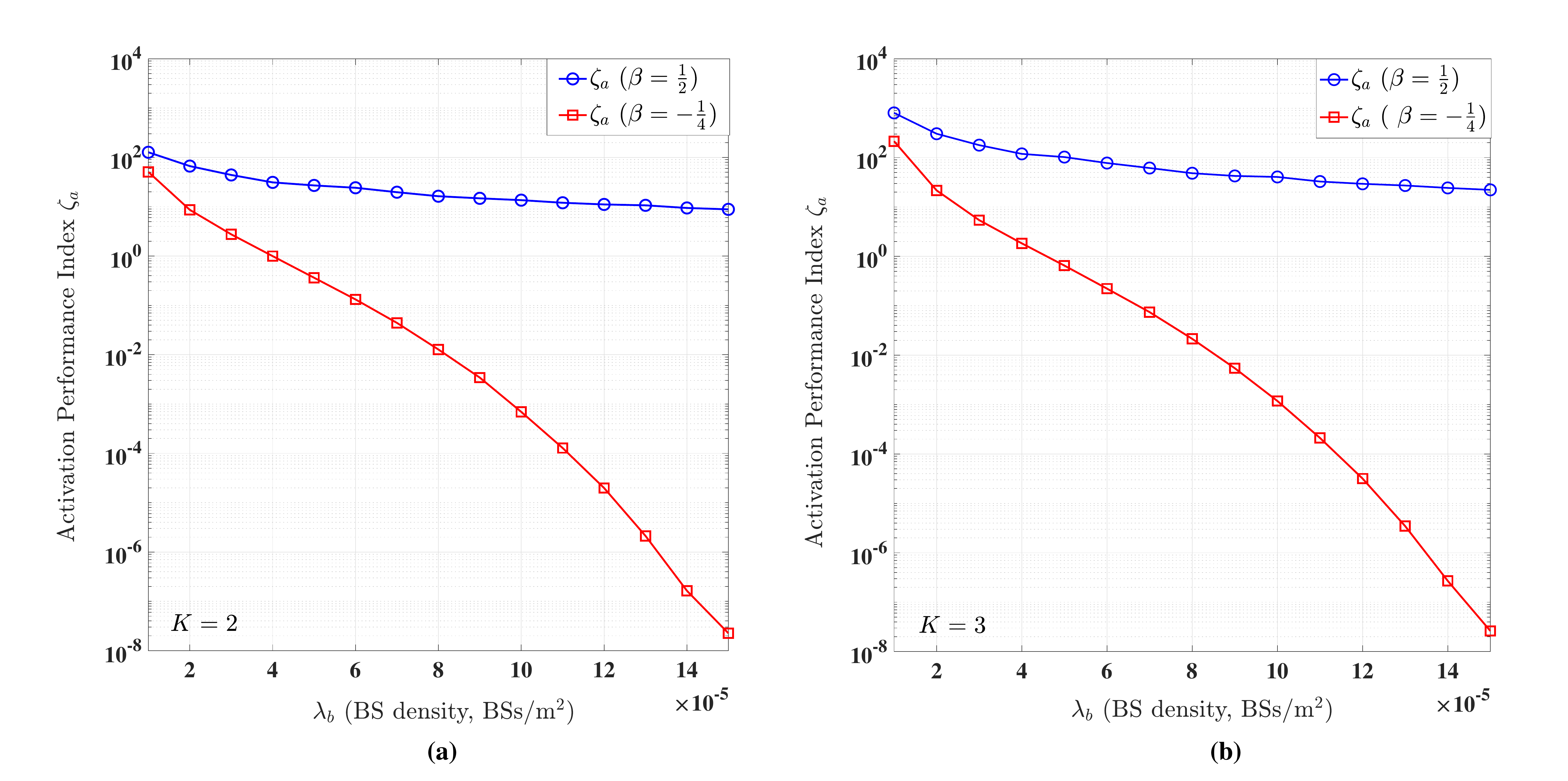}
	\caption{Simulation results of the activation performance index $\zeta_a$ for the proposed downlink power control and BS coordination scheme: (a) $\zeta_a$ versus $\lambda_b$ for the case of $K=2$, (b) $\zeta_a$ versus $\lambda_b$ for the case of $K=3$.}
	\label{Fig:ActPerIndex}
\end{figure*}

The simulation results of the uplink coverage for uplink power control with different values of $\nu$, downlink power control with $\beta=\frac{1}{2}$ and BS coordination with $K=3$ are shown in Fig. \ref{Fig:UplinkCovProbDowPowCon}. As we can see in the figure, all simulation results also pretty close to their corresponding analytical results. Also, we can see that all uplink coverage probabilities are significantly improved if they are compared with the simulation results in Fig. \ref{Fig:UplinkCovProb}. This is because the proposed downlink power control and BS coordination scheme reduces the total activation probability $\eta_a$ so that the interference in the simulation case of Fig. \ref{Fig:UplinkCovProbDowPowCon} is much smaller than that in the simulation case of Fig. \ref{Fig:UplinkCovProb}. Although Fig. \ref{Fig:ActiveProbPowCon_3BSCoor} shows that the false, true and total activation probabilities for the downlink power control with $\beta=\frac{1}{2}$ are much smaller than those for the downlink power control with $\beta=-\frac{1}{4}$, the downlink power control with $\beta=\frac{1}{2}$ actually outperforms that with $\beta=-\frac{1}{4}$ from the perspective of the activation performance index $\zeta_a$ defined in \eqref{Eqn:ActPerIndex}. Fig. \ref{Fig:ActPerIndex} shows the simulation results of the activation performance index $\zeta_a$ for the downlink power control and BS coordination scheme with parameters $\beta\in\{\frac{1}{2},-\frac{1}{4}\}$ and $K\in\{2,3\}$. As shown in the figure, the downlink power control with $\beta=\frac{1}{2}$ is much better than that with $\beta=-\frac{1}{4}$ in terms of $\eta_a$. As $K$ increases form 2 to 3, $\eta_a$ remarkably increases. In addition, $\eta_a$ for the case of $\beta=\frac{1}{2}$ does not vary along $\lambda_b$ as much as $\eta_a$ for the case of $\beta=\frac{1}{2}$. Hence, choosing $\beta=\frac{1}{2}$ is much better than setting $\beta=-\frac{1}{4}$ for the proposed downlink power control in that it attains the much higher performances of activating devices and saving energy. We can use $\eta_a$ to find the best value of $\beta$ for the proposed downlink power control. 

\section{Conclusion}\label{Sec:Conclusion} 
Managing the operating modes of IoT devices through BSs is a paramount task since it helps prolong the working lifetime of the devices in a cellular network that are usually powered by capacity-limited batteries. The goal of this paper is to provide insights into how to make the devices operate in an energy-efficient fashion and transmit in an interference-limited environment. To fulfill this goal, we focus on the fundamental analyses of the statistical properties of the activation signaling process generated by the active BSs in the network. We first define the false, true and total activation probabilities and derive their explicit expressions that reveal downlink power control and BS coordination is an effective means to improve the activation performance. We then propose an energy-efficient uplink power control scheme and derive the neat expression of the uplink coverage probability to show its capability of  saving power and achieving high uplink coverage. An energy-efficient downlink power control and BS coordination scheme is proposed and it is analytically and numerically validated to be able to save the transmit power of the active BSs, reduce the uplink interference from the activated devices and improve the activation performance.

\appendix [Proofs of Propositions]
\numberwithin{equation}{section}
\setcounter{equation}{0}

\subsection{Proof of Proposition \ref{Prop:LapWakeUpSignal}}\label{App:LapTranWakeUpProc}

According to the definition of $W$ in \eqref{Eqn:ShotWakeUpPower} and the conservation property of a PPP in Theorem~1 in \cite{CHLLCW16}, we know $W$ can be alternatively expressed as
\begin{align*}
W =\sum_{i:X_i\in\Phi_b}\omega_i \overline{P}H_i\|X_i\|^{-\alpha}\stackrel{d}{=} \sum_{i:\widehat{X}_i\in\widehat{\Phi}_b}\|\widehat{X}_i\|^{-\alpha},
\end{align*}
where $\stackrel{d}{=}$ stands for the equivalence in distribution, $\widehat{\Phi}_b\defn\{\widehat{X}_i\in\mathbb{R}^2: \widehat{X}_i=(\omega_i\overline{P}H_i)^{-\frac{1}{\alpha}}X_i, X_i\in\Phi_b, i\in\mathbb{N}_+\}$ is a homogeneous PPP of intensity $\lambda_b\mathbb{E}[(\omega \overline{P}H)^{\frac{2}{\alpha}}]$. Thus, the Laplace transform of $W$ can be found as
\begin{align}
\mathcal{L}_{W}(s) &=\mathbb{E}\left[\exp\left(-s\sum_{i:\widehat{X}_i\in\widehat{\Phi}_b}\|\widehat{X}_i\|^{-\alpha}\right)\right]\nonumber\\
&=\mathbb{E}_{\widehat{\Phi}_b}\left[\prod_{i:\widehat{X}_i\in\widehat{\Phi}_b}\exp\left(-s\|\widehat{X}_i\|^{-\alpha}\right)\right]\nonumber\\
&\stackrel{(a)}{=}\exp\left[-\pi\lambda_b\mathbb{E}\left[(wPH)^{\frac{2}{\alpha}}\right]\int_0^{\infty}\left(1-e^{-sr^{-\frac{\alpha}{2}}}\right)\dif r \right]\nonumber\\
&\stackrel{(b)}{=}\exp\left[-\pi\lambda_b\mu (sP)^{\frac{2}{\alpha}}\mathbb{E}\left[H^{\frac{2}{\alpha}}\right]\Gamma\left(1-\frac{2}{\alpha}\right)\right],\label{Eqn:ProofLapTranActSigProc}
\end{align}
where $(a)$ follows from the probability generating functional (PGFL) of a homogeneous PPP and $(b)$ is due to $\mathbb{E}[(\omega \overline{P}H)^{\frac{2}{\alpha}}]=\mu (sP)^{\frac{2}{\alpha}}\mathbb{E}[H^{\frac{2}{\alpha}}]$. Hence, \eqref{Eqn:LapTranWakeUpProc} is obtained because $\mathbb{E}[H^{\frac{2}{\alpha}}]=\Gamma(m+\frac{2}{\alpha})/\Gamma(m)$.

Since $X_1$ is the nearest BS in $\Phi_b$ to the typical device, we know $\|X_{i+1}\|^2=\|X_1\|^2+\|X_i\|^2$ for $i>1$ if $X_i$ the $i$th nearest BS in $\Phi_b$ to the typical device, $\|X_1\|^2\sim\exp(\pi\lambda_b)$ is an exponential RV with mean $\pi\lambda_b$, and $\|X_i\|^2\sim\text{Gamma}(i,\pi\lambda_b)$ \cite{MH12}. Therefore, $W$ in \eqref{Eqn:ShotWakeUpPower} can be equivalently expressed as \cite{CHLLCW16,CHLCSH19}
\begin{align*}
W \stackrel{d}{=} \frac{\omega_1\overline{P}H_1}{\|X_1\|^{\alpha}}+\sum_{j:\widetilde{X}_j\in\widetilde{\Phi}_b}\frac{\omega_j\overline{P}H_j}{(\|X_1\|^2+\|\widetilde{X}_j\|^2)^{\frac{\alpha}{2}}},
\end{align*}
where $\widetilde{\Phi}_b\defn\{\widetilde{X}_j\in\mathbb{R}^2:j\in\mathbb{N}_+\}$ is a homogeneous PPP of intensity $\lambda_b$. For given $\omega_1=0$ and $\|X_1\|^2=x$, $\mathcal{L}_{W|\omega_1=0}(s)=\mathcal{L}_{I_1}(s)$ can be expressed as follows:
\begin{align}
&\mathcal{L}_{I_1}(s) = \mathbb{E}\left[\exp\left(-\sum_{j:\widetilde{X}_j\in\widetilde{\Phi}_b}\frac{s\omega_j\overline{P}H_j}{(x+\|\widetilde{X}_j\|^2)^{\frac{\alpha}{2}}}\right)\right]\nonumber
\end{align}
\begin{align}
&=\mathbb{E}_{\widetilde{\Phi}_b}\left\{\prod_{j:\widetilde{X}_j\in\widetilde{\Phi}_b}\mathbb{E}\left[\exp\left(-\frac{s\omega_j\overline{P}H_j}{(x+\|\widetilde{X}_j\|^2)^{\frac{\alpha}{2}}}\right)\right]\right\} \nonumber\\
&\stackrel{(c)}{=}\exp\left\{-\pi\lambda_b\mu\int_0^{\infty}\left(1-\mathbb{E}\left[\exp\left(-\frac{s\overline{P}H}{(x+r)^{\frac{\alpha}{2}}}\right)\right]\right) \dif r \right\}, \label{Eqn:ProofLapTrans1}
\end{align}
where $(c)$ is obtained by applying the probability generating functional (PGFL) of a homogeneous PPP to $\widehat{\Phi}_b$. By letting $R\sim\exp(1)$, the integral in \eqref{Eqn:ProofLapTrans1} can be alternatively expressed as
\begin{align}
&\int_0^{\infty}\left(1-\mathbb{E}\left[e^{-\frac{s\overline{P}H}{(x+r)^{\frac{\alpha}{2}}}}\right]\right) \dif r =\int_0^{\infty} \mathbb{P}\left[R\leq \frac{s\overline{P}H}{(x+r)^{\frac{\alpha}{2}}}\right]\dif r\nonumber\\
&=\int_x^{\infty} \mathbb{P}\left[u\leq \left(\frac{s\overline{P}H}{R}\right)^{\frac{2}{\alpha}}\right]\dif u =\int_0^{\infty} \mathbb{P}\left[u\leq \left(\frac{s\overline{P}H}{R}\right)^{\frac{2}{\alpha}}\right]\nonumber\\
&\dif u-\mathbb{E}\left\{(s\overline{P})^{\frac{2}{\alpha}}\int_0^{(s\overline{P})^{-\frac{2}{\alpha}}x} \mathbb{P}\left[t\leq \left(\frac{H}{R}\right)^{\frac{2}{\alpha}}\right]\dif t\right\}\nonumber\\
=& \mathbb{E}\bigg\{(s\overline{P})^{\frac{2}{\alpha}} \mathbb{E}\left[H^{\frac{2}{\alpha}}\right]\mathbb{E}\left[R^{-\frac{2}{\alpha}}\right]\nonumber \\
&-\int_0^x \left[1-\left(\frac{t^{\frac{\alpha}{2}}}{t^{\frac{\alpha}{2}}+ms\overline{P}}\right)^m\right]\dif t\bigg\}, \label{Eqn:ProofLapTran2}
\end{align}
where \eqref{Eqn:ProofLapTran2} is equal to \eqref{Eqn:FunInterference} since $\mathbb{E}[H^{\frac{2}{\alpha}}]\mathbb{E}[R^{-\frac{2}{\alpha}}]=\frac{\Gamma(m+\frac{2}{\alpha})}{\Gamma(m)}\Gamma(1-\frac{2}{\alpha})$. Substituting \eqref{Eqn:ProofLapTran2} into \eqref{Eqn:ProofLapTrans1} and averaging over $\|X_1\|^2$ yield the result in \eqref{Eqn:CondLapWd1}. Moreover, we know the following
\begin{align*}
\mathcal{L}_{W}(s)& = \mathbb{P}[\omega_1=0]\mathcal{L}_{W|\omega_1=0}(s)+\mathbb{P}[\omega_1=1]\mathcal{L}_{W_d|\omega_1=1}(s)\\
&= (1-\mu) \mathcal{L}_{I_1}(s)+\mu\mathcal{L}_{D_1+I_1}(s),
\end{align*}
which implies the result in \eqref{Eqn:CondLapWd2} since $\mathbb{P}[\omega_1=1]=\mu$ and $\mathbb{P}[\omega_1=0]=1-\mu$. Finally, note $\mathfrak{I}(0,s\overline{P})=(s\overline{P})^{\frac{2}{\alpha}}\frac{\Gamma(m+\frac{2}{\alpha})}{\Gamma(m)}\Gamma(1-\frac{2}{\alpha})$ based on \eqref{Eqn:FunInterference} and we thus get $\mathcal{L}_W(s)=\exp\{-\pi\lambda_b\mu\mathfrak{I}(0,s\overline{P})\}$ in \eqref{Eqn:LapTranWakeUpProc}, which completes the proof.

\subsection{Proof of Proposition \ref{Prop:ActProb}}\label{App:ProofActProb}
According to \eqref{Eqn:ShotWakeUpPower} and \eqref{Eqn:DefActProb}, the false activation probability can be explicitly written as
\begin{align*}
q_a &= \mathbb{P}[W\geq \theta_a|\omega_0]=1-\mathbb{P}[I_1\leq \theta_a] \\
&= 1-\mathcal{L}^{-1}\left\{\frac{1}{s}\mathcal{L}_{I_1}(s)\right\}(\theta_a).
\end{align*}
Since $\mathcal{L}^{-1}\{\frac{1}{s}F(s)\}(a)=\int_0^a\mathcal{L}^{-1}\{F(s)\}(\tau)\dif \tau$ for $a>0$, the above result of $q_a$ can be expressed as
\begin{align*}
q_a=&1-\int_0^{\theta_a}\mathcal{L}^{-1}\left\{\mathcal{L}_{I_1}(s)\right\}(\tau)\dif\tau\stackrel{(a)}{=}1- \int_0^{\theta_a} \bigg\{\int_0^{\infty} \pi\lambda_b\\
&  \times \exp\bigg\{-\pi\lambda_b \left[x+\mu \mathfrak{I}\left(x,s\overline{P}\right) \right]\bigg\}\dif x\bigg\}(\tau)\dif\tau,
\end{align*}
where $(a)$ follows from the expression of $\mathcal{L}_{I_1}(s)$ in \eqref{Eqn:CondLapWd1}. Thus, the result in \eqref{Eqn:FalseActProb} is obtained. Next, we know $p_a$ can be written as 
\begin{align}
p_a &=\mathbb{P}[W\geq \theta_a|\omega_1]=1-\mathbb{P}\left[D_1+I_1\leq \theta_a\right]\nonumber\\
&=1-\mathcal{L}^{-1}\left\{\frac{1}{s}\mathcal{L}_{D_1+I_1}(s)\right\}\left(\theta_a\right)\label{Eqn:ProofTrueActProb1}
\end{align}
based on \eqref{Eqn:DefActProb}. Also, we have
\begin{align}
&\mathcal{L}^{-1}\left\{\frac{1}{s}\mathcal{L}_{D_1+I_1}(s)\right\}\left(\theta_a\right) \nonumber\\
&= \mathcal{L}^{-1}\left\{\frac{1}{\mu s} \left[\mathcal{L}_W(s)-(1-\mu)\mathcal{L}_{I_1}(s)\right]\right\}(\theta_a)\nonumber\\
&\stackrel{(b)}{=}\frac{1}{\mu}\mathcal{L}^{-1}\left\{\frac{1}{s}\exp\left(-\frac{\pi\lambda_b\mu s^{\frac{2}{\alpha}}}{\mathrm{sinc}(2/\alpha)}\mathbb{E}\left[P^{\frac{2}{\alpha}}\right]\right)\right\}(\theta_a)+\frac{(1-\mu)}{\mu}\nonumber\\
&\times\int_0^{\infty} \pi\lambda_b e^{-\pi\lambda_b x} \mathcal{L}^{-1}\left\{\frac{1}{s}e^{-\pi\lambda_b\mu\mathfrak{I}\left(x,s\right)}\right\}(\theta_a)\dif x, \label{Eqn:ProofTrueActProb2}
\end{align}
where $(b)$ follows from $\mathcal{L}_W(s)$ in \eqref{Eqn:LapTranWakeUpProc} and $\mathcal{L}_{I_1}(s)$ in \eqref{Eqn:CondLapWd1}. Then substituting \eqref{Eqn:ProofTrueActProb2} into \eqref{Eqn:ProofTrueActProb1} gives rise to the expression in \eqref{Eqn:TrueActProb}. Finally, the total activation probability can be expressed as
\begin{align}
\eta_a &= 1-\mathbb{P}[W\leq\theta_a] = 1-\mathcal{L}^{-1}\left\{\frac{1}{s}\mathcal{L}_W(s)\right\}(\theta_a)\nonumber\\
&=1-\int_0^{\theta_a}\mathcal{L}^{-1}\left\{\mathcal{L}_W(s)\right\}(\tau) \dif\tau. \label{Eqn:ProofActProb1}
\end{align} 
Thus, substituting the result in \eqref{Eqn:LapTranWakeUpProc} into  \eqref{Eqn:ProofActProb1} yields the result in \eqref{Eqn:ActProb}. 

\subsection{Proof of Proposition \ref{Prop:UplinkCoveProb}}\label{App:ProofUplinkCoveProb}

According to \eqref{Eqn:DefUplinkSIR} and the Slivnyak theorem, $\gamma_1$ can be equivalently evaluated as if $X_1$ were located at the origin and it can thus be equivalently written as
\begin{align*}
\gamma_1\stackrel{d}{=} \frac{Q_1H_1\|X_1\|^{-\alpha}}{\sum_{j:D_j\in\Phi_{d,a}}Q_iH_i\|D_j\|^{-\alpha}}=\frac{Q_1H_1}{\|X_1\|^{\alpha}I_{d,a}},
\end{align*}
where $I_{d,a}\defn \sum_{j:D_j\in\Phi_{d,a}}Q_iH_i\|D_j\|^{-\alpha}$ is the interference from the activated devices in set $\Phi_{d,a}$. By conditioning on $\frac{\|X_1\|^{\alpha}}{Q_1}=x$, then the uplink coverage can be found as 
\begin{align*}
\eta_c &= \mathbb{P}\left[\frac{1}{\theta_c}\geq 
\frac{I_{d,a}\|X_1\|^{\alpha}}{H_1Q_1}\bigg|\frac{\|X_1\|^{\alpha}}{Q_1}=x\right]\\
&=\mathcal{L}^{-1}\left\{\frac{1}{s}\mathbb{E}\left[\mathcal{L}_{I_{d,a}}\left(\frac{sx}{H}\right)\right]\right\}\left(\frac{1}{\theta_c}\right).
\end{align*} 
Furthermore, for any positive real-valued function $\Psi:\mathbb{R}_+\rightarrow \mathbb{R}_+$ and a non-negative RV $Z$, we have the following identity:
\begin{align}
&\mathbb{E}\left\{\frac{1}{s}\exp\left[-\Psi\left(\frac{s}{Z}\right)\right]\right\}= \int_0^{\infty} \frac{1}{s}\exp\left[-\Psi\left(\frac{s}{z}\right)\right] f_Z(z) \dif z \nonumber\\
&= \int_0^{\infty} \exp\left[-\Psi\left(\frac{1}{t}\right)\right] f_Z(st) \dif t,\label{Eqn:ProofLapTranIden}
\end{align}
where $f_Z(\cdot)$ is the PDF of RV $Z$. By using the identity in \eqref{Eqn:ProofLapTranIden}, we  get the following: 
\begin{align*}
&\frac{1}{s}\mathbb{E}\left[\mathcal{L}_{I_{d,a}}\left(\frac{sx}{H}\right)\right]\\
&=\mathbb{E}\left\{\frac{1}{s}\exp\left[-\pi\rho\eta_a\lambda_d\mathfrak{I}(0,1)\mathbb{E}\left[Q^{\frac{2}{\alpha}}\right]\left(\frac{sx}{H}\right)^{\frac{2}{\alpha}}\right]\right\}\\
&=\int_0^{\infty} \exp\left[-\pi\rho\eta_a\lambda_d\mathfrak{I}(0,1)\mathbb{E}\left[Q^{\frac{2}{\alpha}}\right]\left(\frac{x}{t}\right)^{\frac{2}{\alpha}}\right]f_H(st) \dif t,
\end{align*}
where $f_H(st) = \frac{m}{\Gamma(m)}(mst)^{m-1}e^{-mst}$. Therefore, $\eta_c$ for $\|X_1\|^{\alpha}/Q_1=x$ can be obtained by
\begin{align*}
\eta_c =& \mathcal{L}^{-1}\bigg\{s^{m-1}\int_0^{\infty} \exp\left[-\pi\rho\eta_a\lambda_d\mathfrak{I}(0,1)\mathbb{E}\left[Q^{\frac{2}{\alpha}}\right]\left(\frac{x}{t}\right)^{\frac{2}{\alpha}}\right]\\
&\frac{m}{\Gamma(m)}(mt)^{m-1}e^{-smt} \dif t\bigg\}(x)\\
=& \mathcal{L}^{-1}\bigg\{s^{m-1}\mathcal{L}\bigg\{\exp\left[-\pi\rho\eta_a\lambda_d\mathfrak{I}(0,1)\mathbb{E}\left[Q^{\frac{2}{\alpha}}\right]\left(\frac{xm}{\tau}\right)^{\frac{2}{\alpha}}\right]\\
&\frac{\tau^{m-1}}{\Gamma(m)}\bigg\}(s)\bigg\}(x),
\end{align*}
which is exactly the result in \eqref{Eqn:UplinkCoveProb}.

\subsection{Proof of Proposition \ref{Prop:LapKthCoorActProc}} \label{App:LapKthCoorActProc}

We define $\widetilde{\Phi}_b\defn\{\widetilde{X}_i\in\mathbb{R}^2: i\in\mathbb{N}\}$ which is a homogeneous PPP of intensity $\lambda_b$ and we know $\|X_{i+k}\|^2 = \|X_i\|^2+\|X_k\|^2$ if $X_i$ is the $i$th nearest BS in $\Phi_b$ to the typical device based on the explanation in the proof of Proposition \ref{Prop:LapWakeUpSignal} in Appendix \ref{App:LapTranWakeUpProc}. Accordingly, $\Phi_{b,K}\defn\Phi_b\setminus\{X_k\}_{k=1}^K$ can be equivalently expressed as $\Phi_{b,K}\stackrel{d}{=}  \{X_{i+K}\in\Phi_b:  \|X_{i+K}\|^2 \stackrel{d}{=}\|\widetilde{X}_i\|^2+\|X_K\|^2, \widetilde{X}_i\in\widetilde{\Phi}_b, i\in\mathbb{N}_+ \}$ so that
the Laplace transform of $I_K$ in \eqref{Eqn:CoordActProc} can be equivalently expressed as
\begin{align}
\mathcal{L}_{I_K}(s) =& \mathbb{E}\left[\exp\left(-s\sum_{i:X_i\in\Phi_{b,K}} \frac{\omega_iP_iH_i}{\|X_i\|^{\alpha}}\right)\right]\nonumber\\
=& \mathbb{E}\left[\exp\left(-s\sum_{i:\widetilde{X}_i\in\widetilde{\Phi}_b} \frac{\omega_iP_iH_i}{(\|X_K\|^2+\|\widetilde{X}_i\|^2)^{\frac{\alpha}{2}}}\right)\right]\nonumber
\end{align}
\begin{align}
=& \mathbb{E}_{}\left[\prod_{j:\widetilde{X}_i\in\widetilde{\Phi}_b} \exp\left(- \frac{s\omega_iP_iH_i}{(\|X_K\|^2+\|\widetilde{X}_i\|^2)^{\frac{\alpha}{2}}}\right)\right]\nonumber\\
\stackrel{(a)}{=}&\mathbb{E}_{\|X_K\|^2}\bigg[ \exp\bigg\{-\pi\lambda_b\mu\times \nonumber\\
&\int_{\|X_K\|^2}^{\infty}\mathbb{E}_{PH}\left[1-\exp\left(-\frac{sPH}{t^{\frac{\alpha}{2}}}\right)\right]\dif t\bigg\}\bigg], \label{Eqn:ProofLapTranI_K}
\end{align}
where $(a)$ is obtained by considering that the independence among all $\omega_i$'s is still well preserved after local BS coordination and applying the PGFL of a homogeneous PPP on $\widetilde{\Phi}_b$. By letting $R\sim\exp(1)$, the integral in \eqref{Eqn:ProofLapTranI_K} for a given $\|X_K\|^2=x$ can be further simplified as
\begin{align*}
&\int_x^{\infty}\mathbb{E}_{PH}\left[1-\exp\left(-\frac{sPH}{t^{\frac{\alpha}{2}}}\right)\right]\dif t =\int_x^{\infty} \mathbb{P}\left[R\leq \frac{sPH}{t^{\frac{\alpha}{2}}}\right]\dif t \\
&=\int_x^{\infty} \mathbb{P}\left[t\leq \left(\frac{sPH}{R}\right)^{\frac{2}{\alpha}}\right]\dif t\stackrel{(b)}{=}\mathfrak{I}(x,sP)
\end{align*}
where $(b)$ is obtained based on the results in \eqref{Eqn:ProofLapTran2} in Appendix \ref{App:LapTranWakeUpProc} and the definition of $\mathfrak{I}(x,sP)$ in \eqref{Eqn:FunInterference}. According to \eqref{Eqn:IdenJfun}, we can infer that
\begin{align*}
\mathbb{E}\left[\mathfrak{I}(x,sP)\right] =\mathfrak{I}(0,s) \mathbb{E}\left[P^{\frac{2}{\alpha}}\right]- \epsilon_K x,
\end{align*}
where $\mathbb{E}[P^{\frac{2}{\alpha}}]$ is shown in \eqref{Eqn:FracMomDowlinkPower} and it is found by using \eqref{Eqn:DownLinkPowCon} with $\|X_1\|^2\sim\exp(\pi\lambda_b)$.
Therefore, we have
\begin{align}
\mathcal{L}_{I_K}(s) = \mathbb{E}_{\|X_K\|^2}\bigg[ &\exp\bigg\{-\pi\lambda_b\mu \big(\mathfrak{I}(0,s) \mathbb{E}\left[P^{\frac{2}{\alpha}}\right] \nonumber \\
&- \epsilon_K \|X_K\|^2\big)\bigg\}\bigg], \label{Eqn:ProofLapTranI_K2}
\end{align}
which can be approximated by the result in \eqref{Eqn:LapTranI_K} because $\|X_K\|^2\sim\text{Gamma}(K,\pi\lambda_b)$ and $\epsilon_K$ is not sensitive to $\|X_K\|^2$.

According to \eqref{Eqn:CoordActProc} and the aforementioned results, the Laplace transform of $D_K+I_K$ can be rewritten as follows:
\begin{align*}
&\mathcal{L}_{D_K+I_K}(s) = \mathbb{E}\bigg[\exp\bigg(-s\sum_{k=1}^{K}P_kH_k\|X_k\|^{-\alpha}\\
&-s\sum_{i:\widetilde{X}_i\in\widetilde{\Phi}_b}\frac{\omega_iP_iH_i}{(\|X_K\|^2+\|\widetilde{X}_i\|^2)^{\frac{\alpha}{2}}}\bigg)\bigg]=\mathbb{E}_{\|X_K\|^2}\bigg\{\exp\bigg(-\\
& \pi\lambda_b\int_0^{\|X_K\|^2}\left[1-\mathcal{L}_{PH}\left(sr^{-\frac{\alpha}{2}}\right)\right]\dif r\\
&-\pi\lambda_b\mu\int_{\|X_K\|^2}^{\infty}\left[1-\mathcal{L}_{PH}\left(sr^{-\frac{\alpha}{2}}\right)\right]\dif r\bigg)\bigg\}\\
=&\mathbb{E}_{\|X_K\|^2}\bigg\{\exp\bigg(\pi\lambda_b(1-\mu)\int_{\|X_K\|^2}^{\infty} \left[1-\mathcal{L}_{PH}\left(\frac{s}{r^{\frac{\alpha}{2}}}\right)\right]\dif r\\
&-\pi\lambda_b\int_0^{\infty}\left[1-\mathcal{L}_{PH}\left(\frac{s}{r^{\frac{\alpha}{2}}}\right)\right]\dif r\bigg)\bigg\}\\
=& \frac{\exp\left[-\pi\lambda_b\mathbb{E}\left[P^{\frac{2}{\alpha}}\right]\mathfrak{I}(0,s)\right]}{\mathbb{E}\bigg\{\exp\bigg(-\pi\lambda_b(1-\mu)\int_{\|X_K\|^2}^{\infty} \left[1-\mathcal{L}_{PH}\left(\frac{s}{r^{\frac{\alpha}{2}}}\right)\right]\dif r\bigg)\bigg\}}\\
\stackrel{(c)}{=}&\frac{\exp\left[-\pi\lambda_b\mathbb{E}\left[P^{\frac{2}{\alpha}}\right]\mathfrak{I}(0,s)\right]}{\mathbb{E}\left[\exp\left\{-\pi\lambda_b(1-\mu) \left(\mathfrak{I}(0,s) \mathbb{E}\left[P^{\frac{2}{\alpha}}\right]- \epsilon_K \|X_K\|^2\right)\right\}\right]},
\end{align*}
where $(c)$ follows from the results in \eqref{Eqn:ProofLapTranI_K} and \eqref{Eqn:ProofLapTranI_K2}. The above result of $\mathcal{L}_{D_K+I_K}(s)$ can be approximated by  \eqref{Eqn:LapTranD_K} since $\|X_K\|^2\sim\text{Gamma}(K,\pi\lambda_b)$ and $\epsilon_K$ is not sensitive to $\|X_K\|^2$. Furthermore, we know that $\mathcal{L}_{W_{c,K}}(s)$ can be written as
\begin{align*}
\mathcal{L}_{W_K}=&\mathbb{P}[\mu_1=1]\mathbb{E}\left[\exp\left(-sW_K\right)|\mu_1=1\right]\\
&+\mathbb{P}[\mu_1=0]\mathbb{E}\left[\exp\left(-sW_{c,K}\right)|\mu_1=0\right]\\
=&\mu \mathcal{L}_{D_K+I_K}(s)+(1-\mu) \mathcal{L}_{I_K}(s),
\end{align*}
which is approximately equal to the result in \eqref{Eqn:LapTransWcK} owing to \eqref{Eqn:LapTranI_K} and \eqref{Eqn:LapTranD_K}.


\bibliographystyle{IEEETRAN}
\bibliography{IEEEabrv,Ref_EneEffIoTcellular} 

\end{document}